\documentclass[aps,pra,twocolumn,showkeys,amssymb,floatfix,longbibliography,superscriptaddress]{revtex4-1}

\usepackage{hyperref}
\usepackage{braket}
\usepackage{mathtools}
\usepackage{placeins}
\usepackage{color}
\usepackage{amsmath}
\usepackage{upgreek}
\DeclareRobustCommand\id{\leavevmode\hbox{\small1\normalsize\kern-.33em1}}

\newcommand\minp{\mathop{{\min}'}}

\begin{document}

\title{Benchmarking the Quantum Approximate Optimization Algorithm}

\author{Madita Willsch}
\affiliation{Institute for Advanced Simulation, J\"ulich Supercomputing Centre,\\
Forschungszentrum J\"ulich, D-52425 J\"ulich, Germany}
\affiliation{RWTH Aachen University, D-52056 Aachen, Germany}
\author{Dennis Willsch}
\affiliation{Institute for Advanced Simulation, J\"ulich Supercomputing Centre,\\
Forschungszentrum J\"ulich, D-52425 J\"ulich, Germany}
\affiliation{RWTH Aachen University, D-52056 Aachen, Germany}
\author{Fengping Jin}
\affiliation{Institute for Advanced Simulation, J\"ulich Supercomputing Centre,\\
Forschungszentrum J\"ulich, D-52425 J\"ulich, Germany}
\author{Hans De Raedt}
\affiliation{Zernike Institute for Advanced Materials,\\
University of Groningen, Nijenborgh 4, NL-9747 AG Groningen, The Netherlands}
\author{Kristel Michielsen}
\affiliation{Institute for Advanced Simulation, J\"ulich Supercomputing Centre,\\
Forschungszentrum J\"ulich, D-52425 J\"ulich, Germany}
\affiliation{RWTH Aachen University, D-52056 Aachen, Germany}

\date{\today}

\begin{abstract}
The performance of the quantum approximate optimization algorithm is evaluated by
using three different measures: the probability of finding the ground state, the energy expectation value, and a ratio closely related to the approximation ratio.
The set of problem instances studied consists of weighted MaxCut problems
and 2-satisfiability problems. The Ising model representations of the latter possess unique
ground states and highly degenerate first excited states.
The quantum approximate optimization algorithm is executed
on quantum computer simulators and on the IBM Q Experience.
Additionally, data obtained from the D-Wave 2000Q quantum annealer are used for comparison,
and it is found that the D-Wave machine outperforms the quantum approximate optimization algorithm executed on a simulator. The overall performance of the quantum approximate optimization
algorithm is found to strongly depend on the problem instance.

\end{abstract}

\keywords{quantum computation; quantum annealing; optimization problems; QAOA}

\maketitle

\section{Introduction}
The Quantum Approximate Optimization Algorithm (QAOA) is a variational
method for solving combinatorial optimization problems on a gate-based quantum computer~\cite{farhi14}.
Generally speaking, combinatorial optimization is the task of
finding, from a finite number of objects, that object which
minimizes a cost function.
Combinatorial optimization finds application in real-world
problems including reducing the cost of supply chains,
vehicle routing, job allocation, and so on.
The QAOA is based on a reformulation of
the combinatorial optimization in terms of finding
an approximation to the ground state of a Hamiltonian
by adopting a specific variational ansatz for the trial wave function.
This ansatz is specified in terms of a gate circuit
and involves $2p$ parameters (see below) which have to be
optimized by running a minimization algorithm on a conventional computer.

Alternatively, the QAOA can be viewed as a form of quantum annealing (QA)
using discrete time steps. In the limit that these time steps
become vanishingly small (i.e $p\to\infty$), the adiabatic theorem~\cite{born28} guarantees
that quantum annealing yields the true ground state,
presuming that the adiabatic conditions are satisfied,
thus providing at least one example for which the QAOA yields the correct answer.
In addition, there exists a special class of models for which QAOA with $p=1$ solves the optimization problem exactly~\cite{streif19}.
In general, for finite $p$, there is no guarantee that the QAOA solution
corresponds to the solution of the original combinatorial optimization problem.

Interest in the QAOA has increased dramatically in the past few years
as it may, in contrast to Shor's factoring algorithm~\cite{shor97},
lead to useful results even when used on NISQ devices~\cite{Preskill2018}.
Experiments have already been performed~\cite{otterbach17,Qiang2018}.
Moreover, the field of application, which is optimization, is much larger than, for example, factoring,
rendering the QAOA a possible valuable application for gate-based quantum computers in general.
It has also been proposed to use the QAOA for showing quantum supremacy on near-term devices~\cite{farhi16}.

The aim of this paper is to present a critical assessment of the QAOA, based on results
obtained by simulation, running the QAOA on the IBM Q Experience,
and a comparison with data produced by the D-Wave 2000Q
quantum annealer.

We benchmark the QAOA by applying it to a set of 2-SAT problems with up to 18 variables and weighted MaxCut problems with 16 variables.
We measure performance by means of the energy, a ratio related to the approximation ratio, and the success probability.
We find that the overall success of QAOA depends critically on the problem instance.

The paper is structured as follows:
In Sec.~\ref{sec:background}, we introduce the 2-SAT~\cite{GARE00} and MaxCut~\cite{GARE00} problems
which are used to benchmark the QAOA and review the basic elements
of the QAOA and QA.
Section~\ref{sec:aspects} discusses the procedures
to assess the performance of the QAOA and to compare it with QA.
The results obtained by using simulators, the IBM Q Experience,
and the D-Wave 2000Q quantum annealer are presented in Sec.~\ref{sec:results}.
Section~\ref{sec:conclusion} contains our conclusions.

\section{Theoretical background}\label{sec:background}
\subsection{The 2-SAT Problem}
Solving the 2-satisfiability (2-SAT) problem amounts to finding a true/false
assignment of $N$ Boolean variables such that a given expression is satisfied~\cite{GARE00}.
Such an expression consists of arbitrarily many conjunctions of
clauses that consist of disjunctions of pairs of the Boolean variables (or their
negations), respectively. Neglecting irrelevant constants, problems of this type
can be mapped onto the quantum spin Hamiltonian
\begin{align} H_\mathrm{Ising} = \sum\limits_i h_i\sigma_i^z
+\sum\limits_{\mathclap{i,j}}J_{ij}\sigma_i^z\sigma_j^z,\label{eq:H_ising}
\end{align}
where $\sigma_i^z$ denotes the Pauli $z$-matrix of spin $i$
with eigenvalues $z_i\in\{-1,1\}$.
In the basis that diagonalizes all $\sigma_i^z$ (commonly referred to as the computational basis),
Hamiltonian Eq.~(\ref{eq:H_ising}) is a function of the variables $z_i$.
For the class of 2-SAT problems that we consider, this cost function is integer valued.
Minimizing this cost function answers the question whether there
exists an assignment of $N$ Boolean variables that solves the 2-SAT problem and provides this assignment.

In this paper, we consider a collection of 2-SAT problems that, in terms of Eq.~(\ref{eq:H_ising}),
possess a unique ground state and a highly degenerate first-excited state
and, for the purpose of solving such problems by means of the D-Wave quantum annealer,
allow for a direct mapping onto the Chimera graph~\cite{bunyk14,bian14,Boothby2016}.

\subsection{The MaxCut Problem}
Given an undirected graph $\mathcal{G}$ with vertices $i\in V$ and edges
$(i,j)\in E$, solving the MaxCut problem yields two subsets $S_0$ and
$S_1$ of $V$ such that $S_0\cup S_1 = V$, $S_0 \cap S_1 = \emptyset$, and the
number of edges $(i,j)$ with $i\in S_0$ and $j\in S_1$ is as large as possible~\cite{GARE00}.
In terms of a quantum spin model, the solution
of the MaxCut problem corresponds to the lowest energy eigenstate of the Hamiltonian

\begin{align} H_{\mathrm{MaxCut}} = \sum\limits_{\mathclap{(i,j)\in E}}\sigma_i^z\sigma_j^z,
\label{MAXCUT}
\end{align}
where the eigenvalue $z_i=1$ $(-1)$ of the $\sigma_i^z$ operator
indicates that vertex $i$ belongs to subset $S_0$ ($S_1$).
Clearly, the eigenvalues of Eq.~(\ref{MAXCUT}) are integer valued.

The weighted MaxCut problem is an extension for which the edges $(i,j)$ of the graph $\mathcal{G}$ are
weighted by weights $w_{ij}$.
The corresponding Hamiltonian reads
\begin{align} H_{\mathrm{W}} = \sum\limits_{\mathclap{(i,j)\in
E}}w_{ij}\sigma_i^z\sigma_j^z.\label{eq:H_weighted_maxcut} \end{align}
Obviously, Eq.~(\ref{eq:H_weighted_maxcut}) is a special case of Eq.~(\ref{eq:H_ising}).

\subsection{Quantum Annealing}
Quantum annealing was proposed
as a quantum version of simulated annealing~\cite{finnila94,kadowaki98} and shortly thereafter, the related notion of adiabatic quantum computation has been introduced~\cite{farhi00,childs01}.
The working principle is that an $N$-spin
quantum system is prepared in the state $\ket{+}^{\otimes N}$, which is the
ground state of the initial Hamiltonian $H_\mathrm{init}=-H_0$, where
\begin{align} H_0= \sum\limits_{i=1}^N \sigma_i^x, \end{align}
and $\sigma_i^x$ is the Pauli $x$-matrix for spin $i$.
The Hamiltonian of the system
changes with time according to \begin{align} H(s) = A(s)H_\mathrm{init}
+B(s)H_C,\qquad s=t/t_a, \end{align} where $t_a$ is the total
annealing time, $A(s=0)\gg 1, B(s=0) \approx 0$ and $A(s=1)\approx 0, B(s=1)\gg 1$
(in appropriate units) and $H_C$ is the Hamiltonian corresponding
to the discrete optimization problem (e.g.\ the MaxCut or the 2-SAT problem considered in this paper).

If $|{} \Psi(0) \rangle$ is the ground state of $H_0$,
the adiabatic theorem
says that the state $|{} \Psi(t_a)\rangle$ obtained from the solution of
\begin{align}
i\frac{\partial}{\partial t} |{} \Psi(t)\rangle&=
\left[B(t/t_a) H_C - A(t/t_a) H_0\right] |{} \Psi(t)\rangle, \label{TDSE}
\\
|{}\Psi(0)\rangle&= |{}+ \rangle^{\otimes N}
\;,
\label{ADD1}
\end{align}
with $0\le t\le t_a$, will approach
the ground state (i.e., yield the minimum) of $H_C$ if the variation
of $A(t/t_a)$ and $B(t/t_a)$ is sufficiently smooth and the annealing time $t_a$ becomes infinitely long~\cite{born28}.
Equation~(\ref{TDSE}) is the time-dependent Schr\"odinger equation
with a time-dependent Hamiltonian $B(t/t_a) H_C - A(t/t_a) H_0$.
The formal solution of Eqs.~(\ref{TDSE}) and~(\ref{ADD1}) is given by the time-ordered product of matrix exponentials~\cite{suzuki85}
\begin{gather}
|{} \Psi(t_a)\rangle =
\lim_{p\to\infty}\left\{\prod_{j=1}^p e^{-i\tau(p)\left[B(j/p) H_C - A(j/p) H_0\right]}\right\} |{} + \rangle^{\otimes N}
\nonumber \\
=\lim_{p\to\infty}\left\{\prod_{j=1}^p e^{i\tau(p) A(j/p) H_0}e^{-i\tau(p) B(j/p) H_C} \right\} |{} + \rangle^{\otimes N}
\;,
\label{ADD2}
\end{gather}
where $t=j\tau(p)$ and $\tau(p)=t_a/p$ and we used Trotter's formula~\cite{trotter59} such that $\exp(\tau(H_A+H_B))\to \exp(\tau H_A)\exp(\tau H_B)$ for $\tau\to0$ for two operators $H_A$ and $H_B$.
According to the adiabatic theorem~\cite{born28,ALBA18}, $\lim_{t_a\to\infty}|{} \Psi(t_a)\rangle$ is the ground state of $H_C$.
In practice, quantum annealing is performed with finite $t_a$.

In this paper, we use the D-Wave 2000Q to perform the quantum annealing experiments.

\subsection{Quantum Approximate Optimization Algorithm}
In this section, we briefly review the basic elements of the QAOA~\cite{farhi14}.

Consider an optimization problem for which the objective function is given by
$C(z)=\sum_j C_j(z)$, where $z=z_1z_2\dots z_N$, $z_i\in\{-1,1\}$, and typically, each of
the $C_j(z)$ depends on a few of the $z_i$ only.
If each $C_j(z)$ depends on not more than two of the $z_i$, the mapping of $C(z)$ onto the
Ising Hamiltonian $H_C$ is straightforward.
If $C_j(z)$ depends on the products of three or more $z_i$,
$C(z)$ may still be mapped onto the Ising Hamiltonian,
potentially at the expense of introducing additional auxiliary variables~\cite{Chancellor2017}.
The Ising Hamiltonian is diagonal in the $\sigma^z$ basis,
and the ground state energy, denoted by $E^{(0)}_C$,
corresponds (up to an irrelevant constant) to the minimum of $C(z)$.

The QAOA works as follows.
The quantum computer is prepared in the state $\ket{+}^{\otimes N}$,
i.e.\ the uniform superposition of all computational basis
states, which can be achieved by applying the Hadamard
gates $H^{\otimes N}$ to $\ket{0}^{\otimes N}$.

The next step is to construct a variational ansatz for the wave function according to
\begin{align}
   \ket{\vec\gamma,\vec\beta} = U_B(\beta_p)U_C(\gamma_p)\cdots U_B(\beta_1)U_C(\gamma_1)\ket{+}^{\otimes N},
   \label{KET}
\end{align}
where $\vec\gamma=(\gamma_1,\ldots,\gamma_p)$, $\vec\beta=(\beta_1,\ldots,\beta_p)$ and
\begin{align}
   U_C(\gamma) &= e^{-i\gamma H_C},\\
   U_B(\beta)  &= e^{-i\beta H_0} = e^{-i\beta\sum_{i=1}^N\sigma_i^x}.\label{eq:U_B}
\end{align}
If the eigenvalues of $H_C$ ($H_0$) are integer-valued,
we may restrict the values of the $\gamma_i$ ($\beta_i$) to the interval $[0,2\pi]$ ($[0,\pi]$)~\cite{farhi14}.
In the case of the weighted MaxCut problem (see Eq.~\ref{eq:H_weighted_maxcut}), the $\gamma_i$ cannot be restricted to the interval $[0,2\pi]$, in general.
The parameter $p$ in Eq.~(\ref{KET}) determines the number of independent parameters of the trial state. Modifications of the QAOA also allow for different mixing operators than the one given in Eq.~(\ref{eq:U_B})~\cite{hadfield19}.

As for all variational methods, $\vec\gamma$ and $\vec\beta$ are determined by minimizing the cost function.
In the case at hand, we minimize the expectation value of the Hamiltonian $H_C$, that is
\begin{align}
E_p(\vec\gamma,\vec\beta) = \bra{\vec\gamma,\vec\beta}H_C\ket{\vec\gamma,\vec\beta},
\label{E_p}
\end{align}
as a function of $(\vec\gamma,\vec\beta)$ and denote
\begin{align}
E_p(\vec\gamma^*,\vec\beta^*)=\minp_{\vec\gamma,\vec\beta}E_p(\vec\gamma,\vec\beta),
\end{align}
where $\minp$ denotes a (local) minimum obtained numerically.
In practice, this minimization is carried out on a conventional digital computer.

The quantum computer is prepared in the state $\ket{\vec\gamma,\vec\beta}$ with the current values of $\vec\gamma$ and $\vec\beta$ using the quantum circuit corresponding to Eq.~(\ref{KET}).
According to quantum theory, each measurement of the state of the quantum computer in the computational basis
produces a sample $z$ with probability $P(z)=|\langle z | \vec\gamma,\vec\beta\rangle|^2$.
This procedure is repeated until a sufficiently large number of samples $z$ is collected.
If we want to search for the optimal ($\vec\gamma$, $\vec\beta$) by minimizing
$E_p(\vec\gamma,\vec\beta)$, we can estimate $E_p(\vec\gamma,\vec\beta)$ through
\begin{align}
   E_p(\vec\gamma,\vec\beta) = \sum\limits_z P(z) C(z),
\end{align}
where the sum is over all collected samples $z$ and the probability $P(z)$ is approximated by the relative frequency with which a particular sample $z$ occurs.
When using the quantum computer simulator, the state vector $\ket{\vec\gamma,\vec\beta}$
is known and can be used to compute the matrix element Eq.~(\ref{E_p}) directly, i.e., it is not necessary to produce samples with the simulator.
Obviously, for a complex minimization problem such as Eq.~(\ref{E_p}),
it may be difficult to ascertain that the minimum found is the global minimum.

Once $\vec\gamma^*$ and $\vec\beta^*$ have been determined,
repeated measurement in the computational basis of the state $\ket{\vec\gamma^*,\vec\beta^*}$ of the
quantum computer yields a sample of $z$'s.
In the ideal but exceptional case that $\ket{\vec\gamma^*,\vec\beta^*}$ is the
ground state of $H_C$, the measured $z$ is a representation of that ground state.
In the other case, there is still a chance that the sample contains the ground state. Moreover, one is often not only interested in the ground state  but also in solutions that are close. The QAOA produces such solutions because even if $\ket{\vec\gamma^*,\vec\beta^*}$ is not the ground state, it is likely that $z$'s for which $C(z) \le E_p(\vec\gamma^*,\vec\beta^*)$ are generated.

The QAOA can also be viewed as a finite-$p$ approximation of Eq.~(\ref{ADD2}),
where in addition the constraint
that the coefficients of $H_0$ and $H_C$ derive from the functions $A(j/p)$ and $B(j/p)$ is relaxed.
Instead of $|{} \Psi(t_a)\rangle$, we now have
\begin{eqnarray}
|{} \vec\gamma ,\vec\beta \rangle&=&
\left\{\prod_{j=1}^p e^{-i\beta_j H_0} e^{-i\gamma_j H_C} \right\} |{} + \rangle^{\otimes N}
\;,
\label{ADD3}
\end{eqnarray}
where the $\gamma_j$'s and $\beta_j$'s are to be regarded as parameters that can be chosen freely.
In Appendix~\ref{app:relation}, we show that if $\vec\gamma$ and $\vec\beta$
are chosen according to the linear annealing schedule,
we recover the finite-$p$ description of the quantum annealing process.
The underlying idea of the QAOA
is that even for small $p$, Eq.~(\ref{ADD3})
can be used as a trial wave function in the variational sense.
For finite $p$, the QAOA only differs from other variational
methods of estimating ground state properties~\cite{PERU14,MALL16,KAND17,YANG17,TING19}
by the restriction to wave functions of the form of Eq.~(\ref{ADD3}).

\subsection{Performance Measures}
We consider three measures for the quality of the solution,
namely (M1) the probability for finding the ground state
(called success probability in what follows)
which should be as large as possible,
(M2) the value of $E_p(\vec\gamma^*,\vec\beta^*)$ which should be as
small as possible,
and (M3) the ratio defined by
\begin{align}
   r = \frac{E_p(\vec\gamma^*,\vec\beta^*)-E_{\mathrm{max}}}{E_{\mathrm{min}}-E_{\mathrm{max}}},
   \label{eq:r}
\end{align}
which should be as close to one as possible and indicates how close the
expectation value $E_p(\vec\gamma^*,\vec\beta^*)$ is to the optimum.
For the set of problems treated in this paper, the eigenvalues of the problem
Hamiltonian can take negative and positive values. We denote the smallest and largest
eigenvalues by $E_{\mathrm{min}}$ and $E_{\mathrm{max}}$, respectively.
As a consequence, the ratio $E_p(\vec\gamma^*,\vec\beta^*)/E_\mathrm{min}$ can have
negative and positive values. By subtracting the largest eigenvalue $E_\mathrm{max}$,
we shift the spectrum to be nonpositive and the ratio $r$ is thus nonnegative with $0\le r\le1$.
In computer science, a \emph{$\rho$-approximation algorithm} is a polynomial-time algorithm
which returns for all possible instances of an optimization problem, a solution with
cost value $V$ such that
\begin{align}
\frac{V}{V'} \ge \rho,\label{eq:rho}
\end{align}
where $V'$ is the cost of the optimal solution~\cite{williamson11}.
For randomized algorithms, the \emph{expected cost} of the solution has to be at least $\rho$
times the optimal solution~\cite{Goemans95}.
The constant $\rho$ is called \emph{performance guarantee} or \emph{approximation ratio}.
The ratio $r$ corresponds to the left-hand side of the
definition of the approximation ratio $\rho$ (Eq.~(\ref{eq:rho})). Since we
cannot investigate all possible problem instances, we use $r$ only as a measure for the subset of
instances that we have selected.

We do not consider the run time or the time-to-solution as performance measures since the timing results obtained from the simulator may not be representative for QAOA performed on a real device.
Obtaining a single sample (for $p=1$) on the IBM Q 16 Melbourne processor takes about $3\,\upmu$s.
However, we used the IBM Q Experience for a grid search only.
Usually, the waiting time in the queue is much longer than the run time and we did not perform QAOA with the optimization step on the real device.
However,
we also performed quantum annealing on the D-Wave 2000Q quantum annealer with an annealing time of $t_a=3\,\upmu$s.

\begin{figure}[bt]
   \centering
   \includegraphics[width=0.3\textwidth]{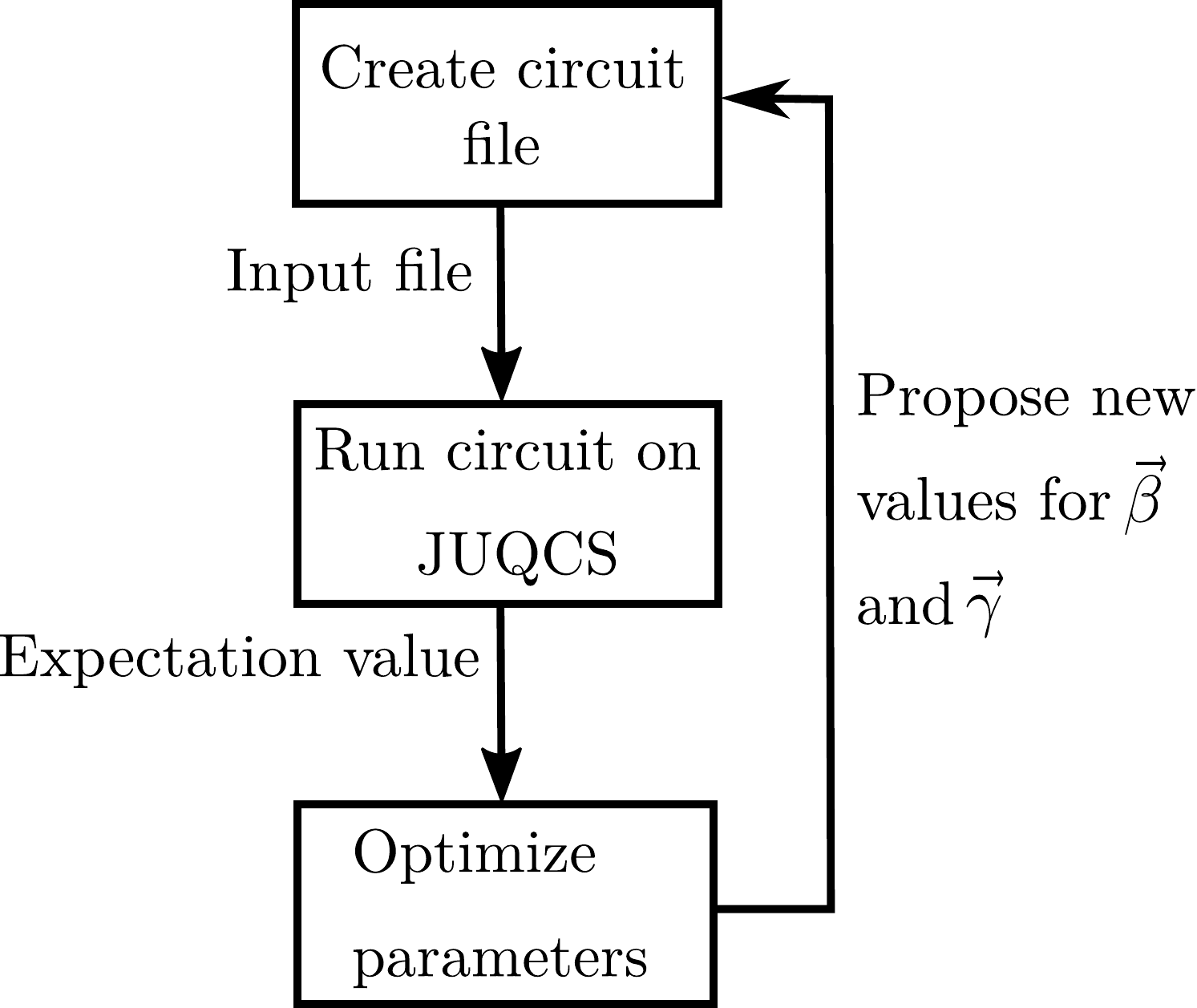}
   \caption{Sketch of the workflow for executing the QAOA with $p>1$.}
   \label{fig:sketch}
\end{figure}
As measures (M1) and (M3) require knowledge of the ground state of $H_C$,
they are only useful in a benchmark setting.
In a real-life setting, only measure (M2) is of practical use.
For the simplest case $p=1$ and a triangle-free (connectivity) graph,
the expectation value of the Hamiltonian
\begin{align}
     H_C = \sum\limits_i h_i\sigma_i^z +\sum\limits_{\mathclap{(i,j)\in E}}J_{ij}\sigma_i^z\sigma_j^z,\label{eq:Hc}
\end{align}
can be calculated analytically.
The result is given by
\begin{align}
   E(\gamma,\beta)&= \sum\limits_i h_i\sin(2\beta)\sin(2\gamma h_i)\prod\limits_{\mathclap{j:(i,j)\in E}}\cos(2\gamma J_{ij})\nonumber\\
   &+\sum\limits_{\mathclap{(i,j)\in E}}J_{ij}\Bigg(\sin^2(2\beta)\sin(2\gamma h_i)\sin(2\gamma h_j) \nonumber\\
   &\quad\times\prod\limits_{\mathclap{k\neq j:(i,k)\in E}}\cos(2\gamma J_{ik})\prod\limits_{\mathclap{l\neq i:(j,l)\in E}} \cos(2\gamma J_{jl})\nonumber\\
     &+\frac{1}{2}\sin(4\beta) \sin(2\gamma J_{ij})\Bigg( \cos(2\gamma h_i)\prod\limits_{\mathclap{k\neq j:(i,k)\in E}} \cos(2\gamma J_{ik})\nonumber\\
     &\quad+\cos(2\gamma h_j)\prod\limits_{\mathclap{l\neq i:(j,l)\in E}} \cos(2\gamma J_{jl}) \Bigg) \Bigg),\label{eq:expv_formula}
\end{align}
where the products are over those vertices that share an edge with the indicated
vertex. For $h_i=0$ and $J_{ij}=1/2$, Eq.~(\ref{eq:expv_formula}) is the same as
Eq.~(15) in Ref.~\cite{wang18}, up to an irrelevant constant contribution.
We use Eq.~(\ref{eq:expv_formula}) as an independent check for our numerical results.

\section{Practical aspects}\label{sec:aspects}
We adopt two different procedures for testing the QAOA.
For $p=1$, we evaluate $E_p(\gamma,\beta)$ for points $(\gamma,\beta)$ on a regular 2D grid.
We create the corresponding gate circuit using Qiskit~\cite{qiskit} and execute it on the IBM simulator
and the IBM Q Experience~\cite{ibmquantumexperience}.
Instances which are executed on the IBM Q Experience natively fit, meaning that they directly map onto the architecture such that no additional SWAP-gates are needed.

For the QAOA with $p>1$, we perform the procedure shown in Fig.~\ref{fig:sketch}.
Given $p$ and values of the parameters $\vec\beta$ and $\vec\gamma$,
a computer program defines the gate circuit in the J\"ulich universal quantum
computer simulator (JUQCS)~\cite{deraedt18} format.
JUQCS executes the circuit and returns the expectation value
of the Hamiltonian $H_C$ in the state $\ket{\vec\gamma,\vec\beta}$ (or the success probability).
This expectation value (or this success probability) in turn is passed to a Nelder-Mead minimizer~\cite{NelderMead1965,numericalrecipes} which
proposes new values for $\vec\beta$ and $\vec\gamma$.
This procedure is repeated until $E_p(\vec\gamma,\vec\beta)$ (or the success probability)
reaches a stationary value.
Obviously, this stationary value does not need to be the global minimum
of $E_p(\vec\gamma,\vec\beta)$ (or the success probability).
In particular, if $E_p(\vec\gamma,\vec\beta)$ (or the success probability as a function of $\vec\gamma$ and $\vec\beta$) has many local minima,
the algorithm is likely to return a local minimum.
This, however, is a problem with minimization in general and is not specific to the QAOA.
In practice, we can only repeat the procedure
with different initial values of $(\vec\gamma,\vec\beta)$
and retain the solution that yields the smallest $E_p(\vec\gamma,\vec\beta)$ (or the highest success probability).
For the 18-variable problems, the execution time of a single cycle, as depicted in Fig.~\ref{fig:sketch}, is less than a second for small $p$ and even for $p\approx 40--50$, the execution of a cycle takes about one second. The execution time of the complete optimization then depends on how many cycles are needed for convergence.

For the QAOA, many (hundreds of) evaluations $N_{ev}$ of $E_p(\vec\gamma,\vec\beta)$ are
necessary for optimizing the parameters $\vec\gamma$ and $\vec\beta$.
A point that should be noted is that we obtain the success
probability for the QAOA from the state vector and that with little effort, we can calculate
$E_p(\vec\gamma,\vec\beta)$ in that state when using the simulator.
In contrast, when using a real quantum device,
in practice $E_p(\vec\gamma,\vec\beta)$ is estimated
from a (small) sample of $N_S$ values of $\langle z|H_C|z\rangle$.
Therefore, using the QAOA on a real device only makes sense if
the product $N_S\cdot N_{ev}$ is much smaller than the dimension of
the Hilbert space of $2^N$. Otherwise the amount of work
is comparable to exhaustive search over the $2^N$ basis states of the Hilbert space.

For the quantum annealing experiments on the D-Wave quantum annealer,
we distribute several copies of the problem instance (that is the Ising Hamiltonian Eq.~(\ref{eq:Hc}))
on the Chimera graph and repeat the annealing procedure to collect statistics
about the success probability and the ratio $r$.
Since we do not need a minor embedding for the problem instances considered,
we can directly put 244 (116, 52) copies of the 8-variable (12-variable, 18-variable,
respectively) instances simultaneously on the D-Wave 2000Q quantum annealer and we only need 250
(500, 1000, respectively) repetitions for proper statistics to infer the success
probability. If we are not interested in estimating the success probability but only need the
ground state to be contained in the sample, much less repetitions are necessary.

\section{Results}\label{sec:results}
\subsection{QAOA with $p=1$}

\begin{figure}[tb]
   \centering
   \includegraphics[width=0.5\textwidth]{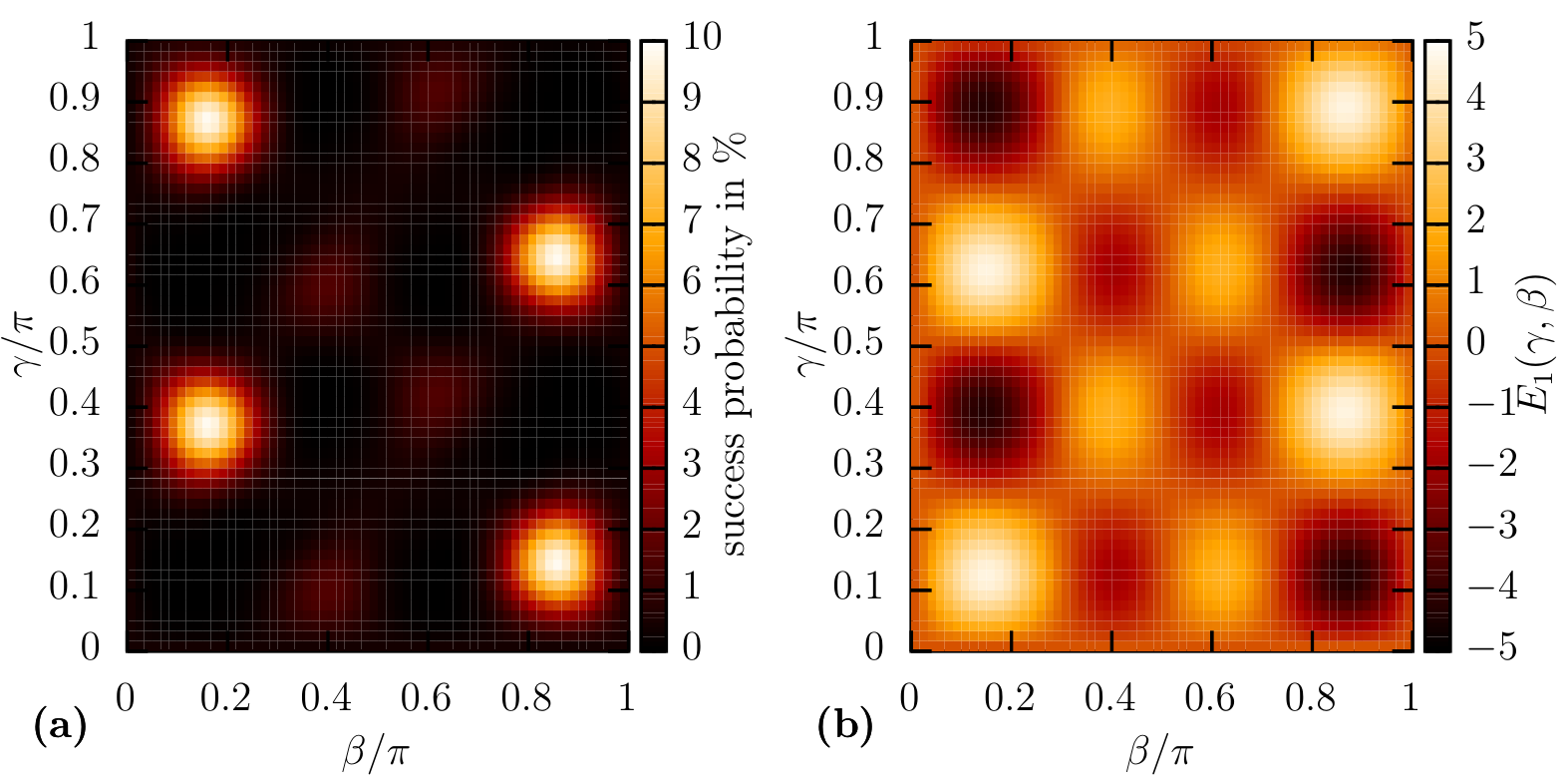}
   \caption{(color online) Simulation results for
   the 8-variable 2-SAT problem instance (A) (see Table~\ref{tab:2sat8} in Appendix~\ref{app:problem_instances}) as a function of $\gamma$ and $\beta$ for $p=1$.
   (a) Success probability (b) $E_1(\gamma,\beta)$.}
   \label{fig:2sat8_both}
\end{figure}

\begin{figure}[tb]
   \centering
   \includegraphics[width=0.5\textwidth]{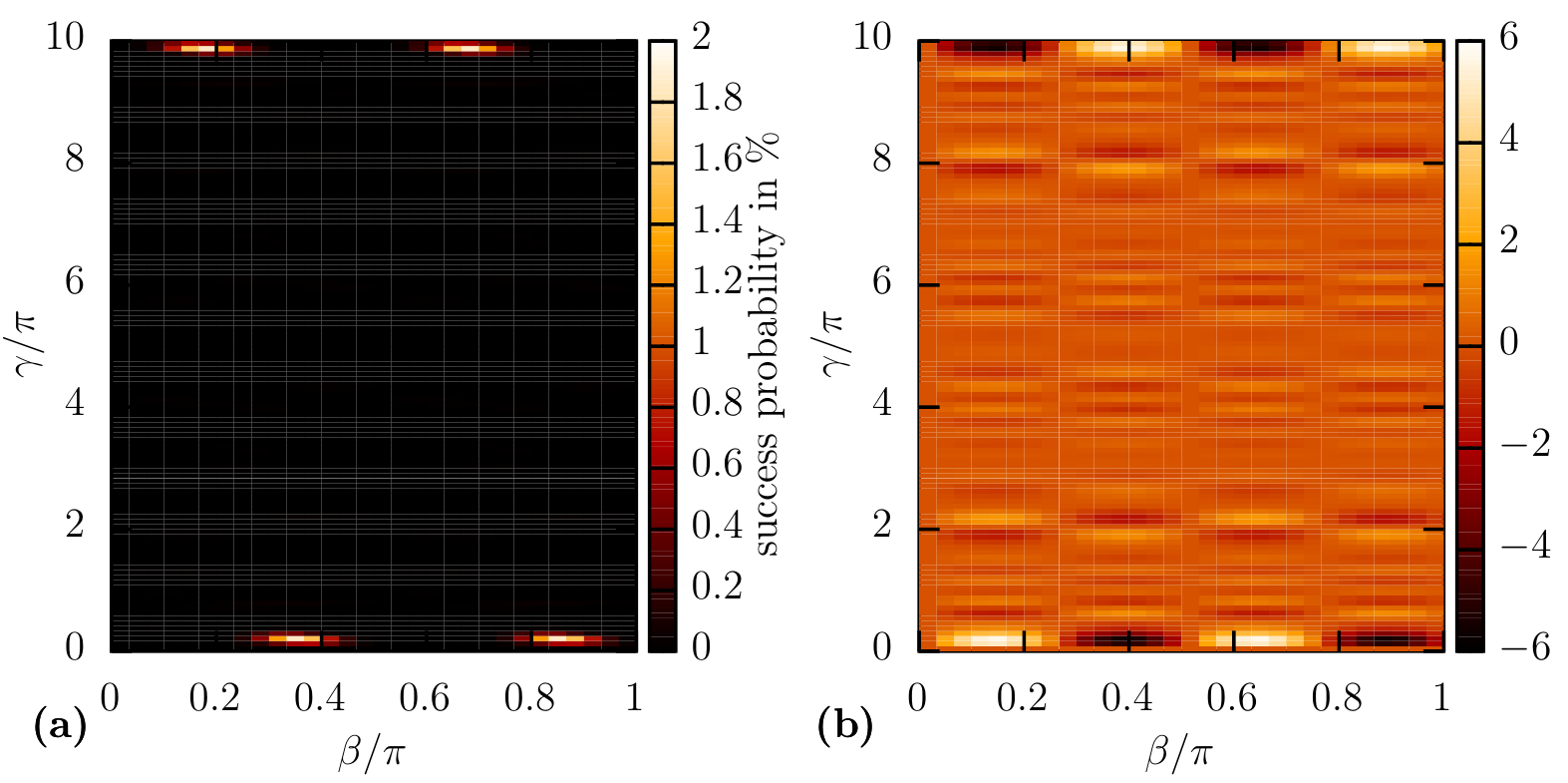}
   \caption{(color online) Same as Fig.~\ref{fig:2sat8_both} except that
   instead of the 8-variable 2-SAT problem instance (A), we solve
   a 16-variable MaxCut problem given in Table~\ref{tab:maxcut} in Appendix~\ref{app:problem_instances}.}
   \label{fig:maxcut_both}
\end{figure}

Figures~\ref{fig:2sat8_both} and \ref{fig:maxcut_both} show the success probability and the expectation value $E_1(\gamma,\beta)$,
i.e., after applying the QAOA for $p=1$, as a function of $\gamma$ and $\beta$
for a 2-SAT problem with 8 spins and
for a 16-variable weighted MaxCut problem, respectively, as obtained by using the IBM Q simulator.
The specifications of the problem instances are given in Appendix~\ref{app:problem_instances}.
With the simulator, the largest success probability that has been obtained
for the 8-variable 2-SAT problem is about 10\% and about 2\% for the 16-variable weighted MaxCut problem.
We find that regions with high success probability correspond to small energy expectation values,
as expected (see Figs.~\ref{fig:2sat8_both} and \ref{fig:maxcut_both}). However,
the values of $(\gamma,\beta)$ for which the success probability is the largest
and $E_1(\gamma,\beta)$ is the smallest differ slightly.


\begin{figure}[tb]
   \centering
   \includegraphics[width=0.5\textwidth]{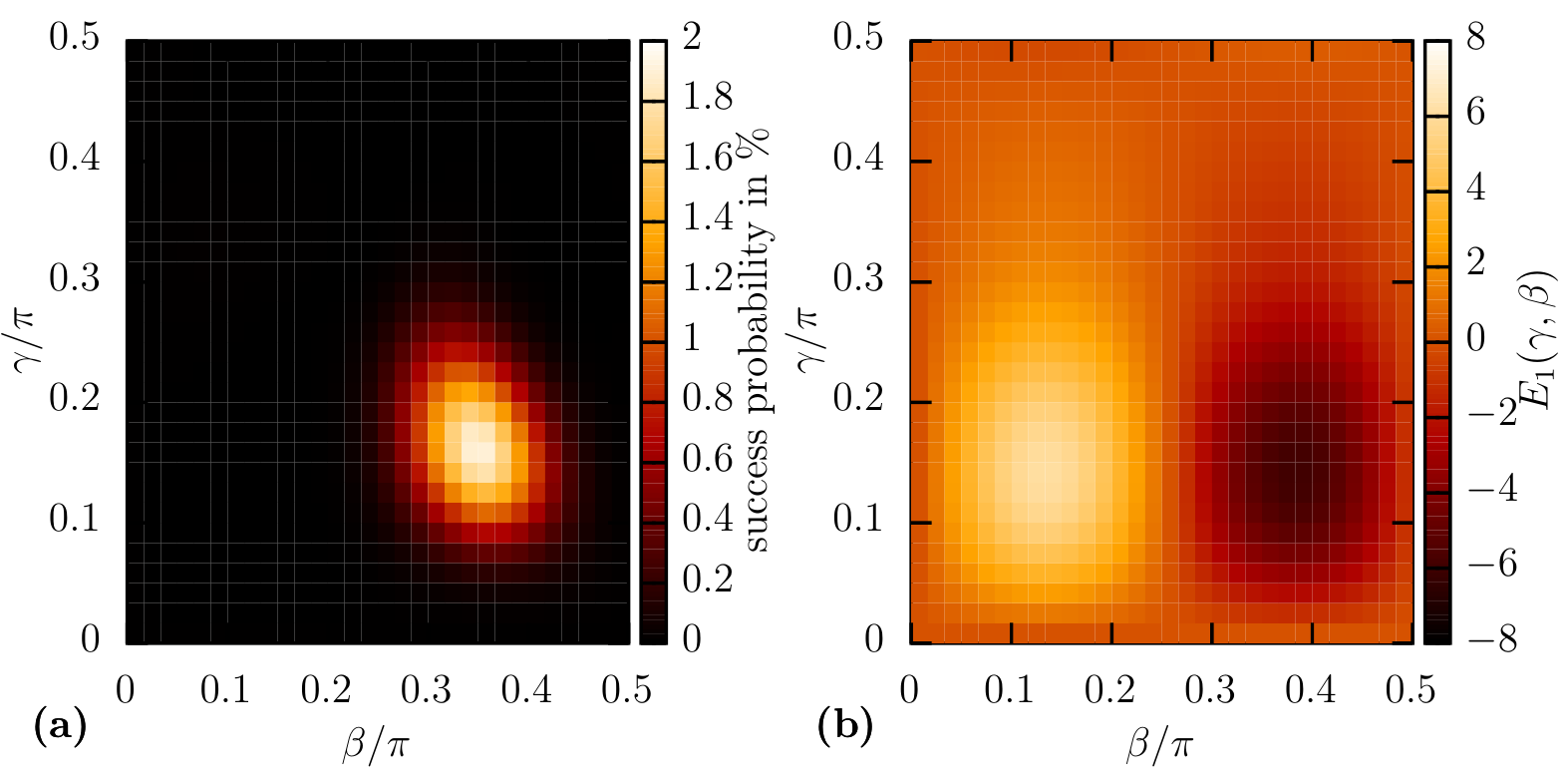}
   \caption{(color online) The same as Fig.~\ref{fig:maxcut_both} except that the part containing
   the maximum success probability is shown on a finer grid.}
   \label{fig:maxcut_both_small}
\end{figure}

As mentioned earlier, if the Hamiltonian Eq.~(\ref{eq:Hc}) does not have integer eigenvalues,
which is the case for the weighted MaxCut problem that we consider (see Eq.~(\ref{eq:H_weighted_maxcut})), the periodicity of
$E_1(\gamma,\beta)$ with respect to $\gamma$ is lost.
Therefore, the search space for $\gamma$ increases severely.
Moreover, the landscape of the expectation value $E_1(\gamma,\beta)$ exhibits many local minima.
Fortunately, for the case at hand, it turns out that the largest success probability can still be
found for $\gamma \in[0,2\pi]$.
Plots with a finer $\gamma$ grid around the largest success probability and the smallest
value of $E_1(\gamma,\beta)$ are shown in
Figs.~\ref{fig:maxcut_both_small}a and~\ref{fig:maxcut_both_small}b, respectively.
Clearly, using a simulator and for $p=1$, it is not difficult to find
the largest success probability or the smallest $E_1(\gamma,\beta)$,
as long as the number of spins is within the range that the simulator can handle.

\begin{figure}[tb]
   \centering
   \includegraphics[width=0.5\textwidth]{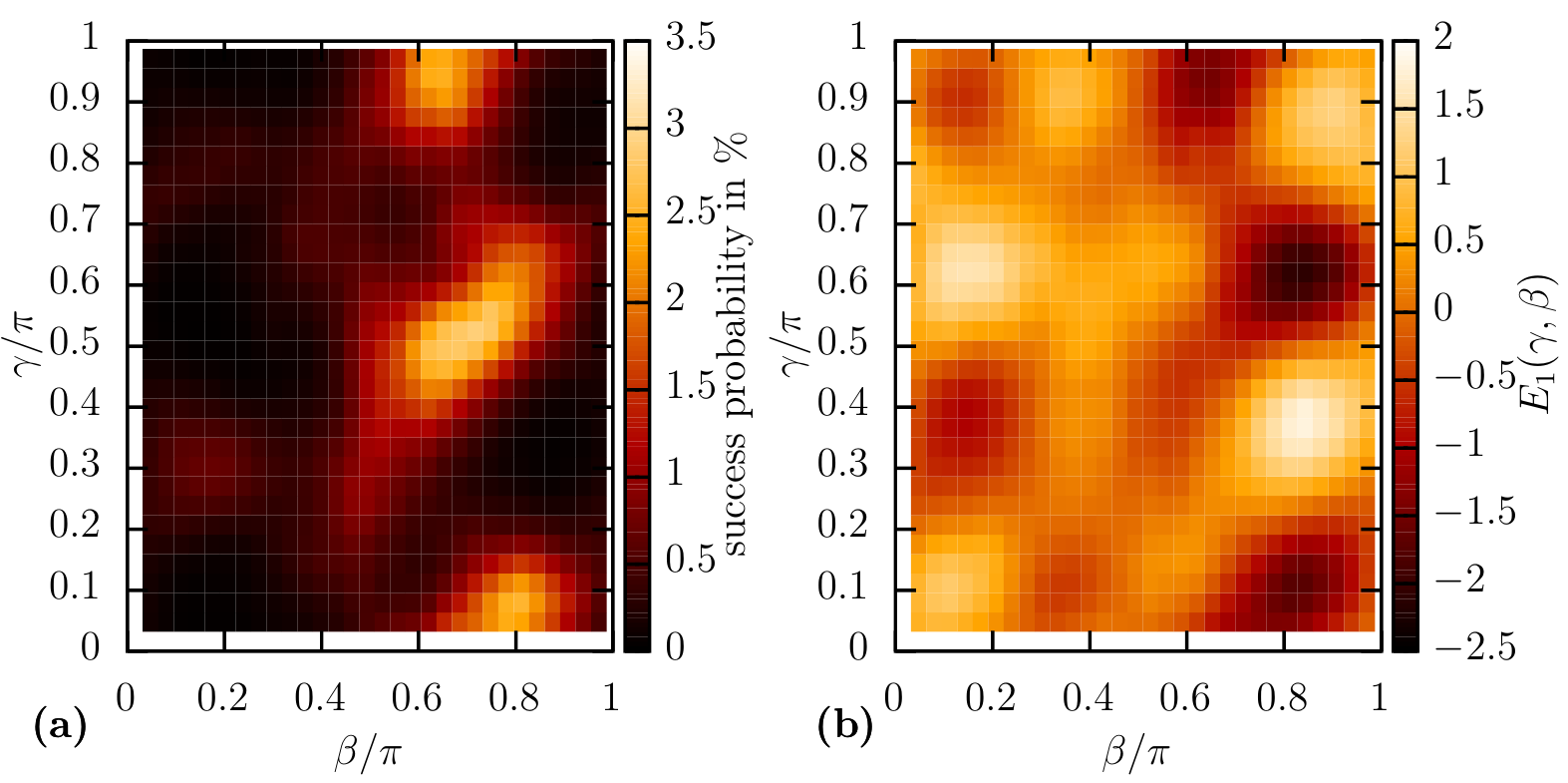}
   \caption{(color online) Same as Fig.~\ref{fig:2sat8_both} except that instead
   of using the IBM Q simulator, the results have been obtained by using the quantum processor IBM Q 16 Melbourne of the IBM Q Experience.}
   \label{fig:2sat8_ibm}
\end{figure}

The results for the same 8-variable 2-SAT problem instance shown in Fig.~\ref{fig:2sat8_both},
but obtained by using the quantum processor IBM Q 16 Melbourne~\cite{ibmquantumexperience}, are shown in Fig.~\ref{fig:2sat8_ibm}.
To obtain an estimate of the success probability, for each pair of $\beta$ and $\gamma$,
we performed seven runs of 8192 samples each.
Note that in this case, the total number of samples per grid point ($57344$) is much
larger than the number of states $2^8=256$.
Thus, we can infer the success probability with very good statistical accuracy.
However, such an estimation is feasible for small system sizes only.
By comparing Figs.~\ref{fig:2sat8_both} and~\ref{fig:2sat8_ibm},
we conclude that the IBM Q Experience results for the success probability do not
bear much resemblance to those obtained by the simulator.
However, the IBM Q Experience results for $E_1(\gamma,\beta)$
show some resemblance to those obtained by the simulator.
It seems that at this stage of hardware development,
real quantum computer devices have serious problems producing data
that are in qualitative agreement with the $p=1$ solution Eq.~(\ref{eq:expv_formula}).

\begin{figure}[tb]
   \centering
   \includegraphics[width=0.45\textwidth]{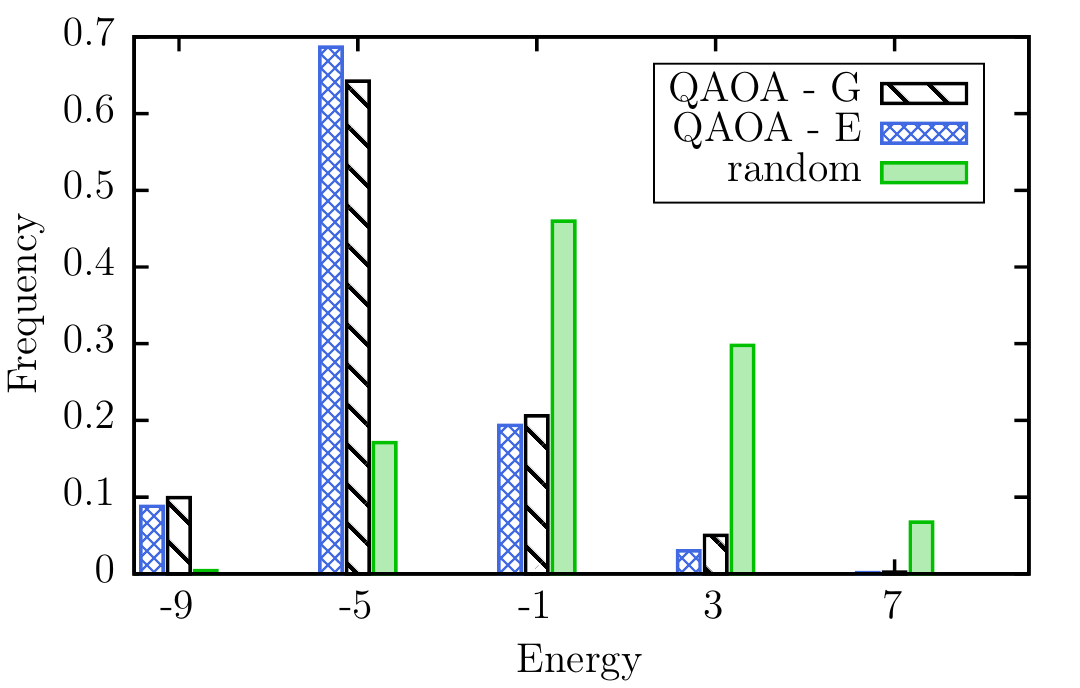}
   \caption{(color online) Frequencies of
   sampled energies $\langle z|H_C|z\rangle$ for the 8-variable 2-SAT problem instance (A),
   obtained by simulation of the QAOA with $p=1$.
   Black (striped): $(\gamma,\beta)$ maximize the success probability;
   blue (squared): $(\gamma,\beta)$ minimize $E_1(\gamma,\beta)$;
   green (solid): $\gamma=\beta=0$ corresponding to random sampling.}
   \label{fig:2sat8_19_sampling}
\end{figure}

\begin{figure}[tb]
   \centering
   \includegraphics[width=0.5\textwidth]{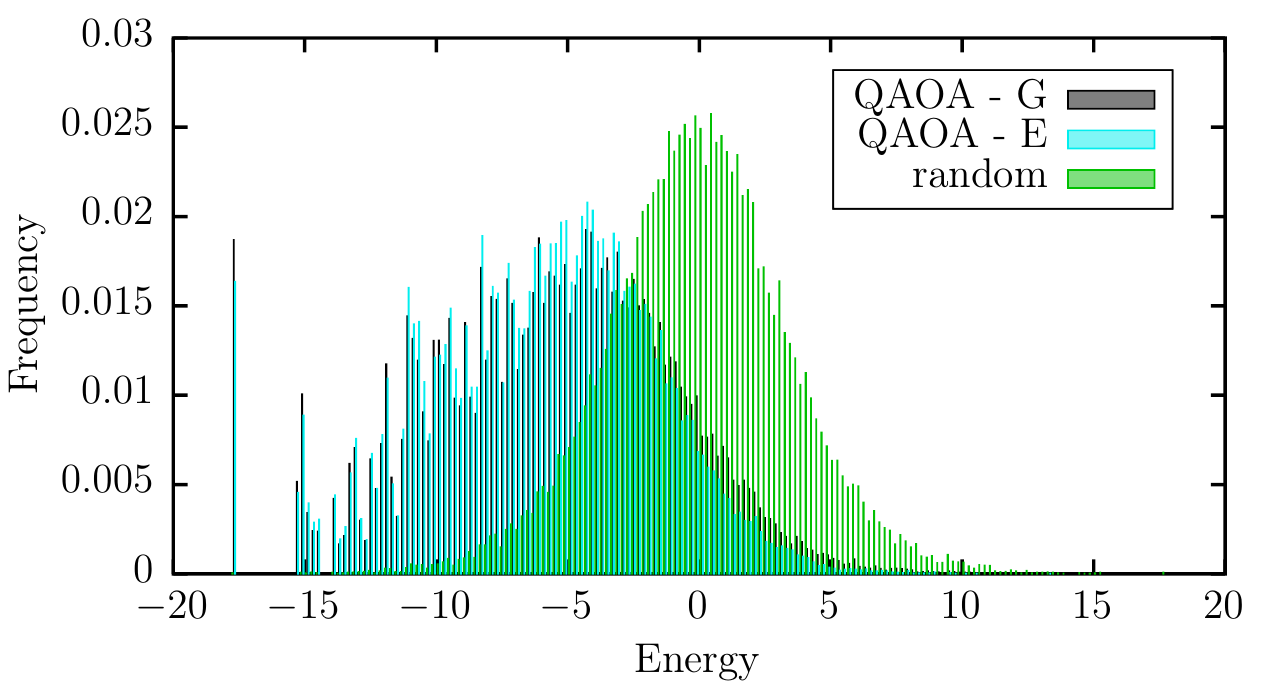}
   \caption{(color online) Same as Fig.~\ref{fig:2sat8_19_sampling} except that
   instead of the 8-variable 2-SAT problem instance (A), we solve
   a 16-variable MaxCut problem.}
   \label{fig:maxcut_sampling}
\end{figure}

Figures~\ref{fig:2sat8_19_sampling} and \ref{fig:maxcut_sampling} show the
distributions of
$\langle z|H_C|z\rangle$ where the states $z$ are samples generated
with probability $|\langle z|\gamma,\beta\rangle|^2$
for the values of $\gamma$ and $\beta$
that maximize the $p=1$ success probability (black, ``QAOA - G'') and minimize $E_1(\gamma,\beta)$ (blue, ``QAOA - E'')
for the 8-variable 2-SAT problem and the 16-variable weighted MaxCut problem, respectively.
For comparison, we also show the corresponding distributions obtained by random sampling (green).
Although for $p=1$, the QAOA enhances the success probability compared to random sampling,
for the 16-variable MaxCut problem, the probability of finding the ground state is less than $2\%$,
as shown in Fig.~\ref{fig:maxcut_sampling}.

From these results, we conclude that as the number of variables increases, the largest
success probability that can be achieved with the QAOA for $p=1$ is rather small.
Moreover, the $p=1$ results obtained on a real gate-based quantum device are
of very poor quality, suggesting that the prospects of performing $p>1$ on such devices
are, for the time being, rather dim.
However, we can still use JUQCS to benchmark the performance of the QAOA for $p>1$ on an ideal quantum computer by adopting the procedure sketched in Fig.~\ref{fig:sketch}. Simulations of the QAOA on noisy quantum devices are studied in Ref.~\cite{guerreschi19}.

\subsection{QAOA for $p>1$}

\begin{figure}[tb]
   \centering
   \includegraphics[width=0.45\textwidth]{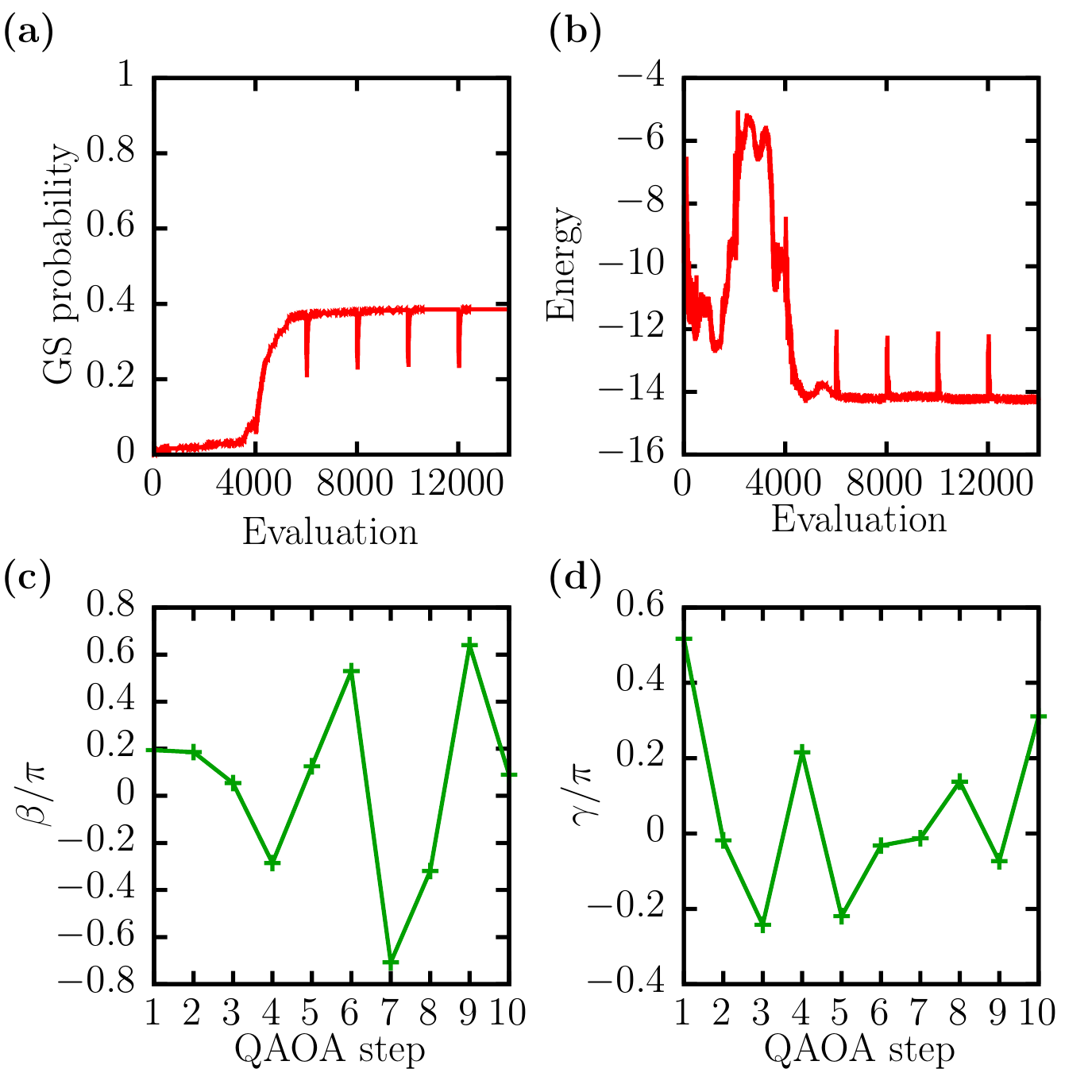}
   \caption{(color online)
   Simulation results of the $p=10$ QAOA applied to the 18-variable 2-SAT problem instance (A) (see Appendix~\ref{app:problem_instances}).
   Shown are (a) the success probability, (b)
   the energy $E_{10}=E_{10}(\vec\gamma,\vec\beta)$
   as a function of the iteration steps of the Nelder-Mead algorithm,
   during the minimization of the success probability,
   (c) the values $\beta_i$ and (d) the values $\gamma_i$ for $i=1,\ldots, 10$
   as obtained after 14000 Nelder-Mead iterations.
   The initial values of $(\vec\gamma,\vec\beta)$ are chosen such that
      $\beta_1=\beta_1^*$ and $\gamma_1=\gamma_1^*$, where $\beta_1^*$ and $\gamma_1^*$
      are the optimal values extracted from the $p=1$ QAOA minimization of the success probability,
      and all other $\beta_i$ and $\gamma_i$ are random. For this 2-SAT problem, the actual ground state energy is $E_C^{(0)}=-19$.}
   \label{fig:2sat_18spin_p2_random}
\end{figure}

Figure~\ref{fig:2sat_18spin_p2_random} shows results
produced by combining JUQCS and the Nelder-Mead algorithm~\cite{NelderMead1965,numericalrecipes}
which demonstrate that for $p=10$ and the 18-variable 2-SAT problem instance (A) (see Appendix~\ref{app:problem_instances}), there exist $\vec\gamma$ and $\vec\beta$ which produce a success probability of roughly 40\%.
The minimization of the success probability starts with values for $(\vec\gamma,\vec\beta)$ which are chosen such that $\gamma_1=\gamma_1'$ and $\beta_1=\beta_1'$, where $\gamma_1'$ and $\beta_1'$ denote the optimal values for the success probability extracted from the $p=1$ QAOA simulation data,
and all other $\gamma_i$ and $\beta_i$ are random.
From Fig.~\ref{fig:2sat_18spin_p2_random}(a), we conclude that
the Nelder-Mead algorithm is effective in finding a minimum of the success probability
(the spikes in the curves correspond to restarts of the search procedure).
As can be seen in Fig.~\ref{fig:2sat_18spin_p2_random}(b), the energy expectation $E_{p=10}$
also converges to a stationary value as the number of Nelder-Mead iterations increases.
The values of $\beta_i$ and $\gamma_i$ at the end of the minimization process
are shown in Fig.~\ref{fig:2sat_18spin_p2_random}(c,d).

Note that the use of the success probability as the cost function to be minimized
requires the knowledge of the ground state, i.e.\ of the solution of the optimization problem.
Obviously, for any problem of practical value, this knowledge is not available but
for the purpose of this paper, that is, for benchmarking purposes,
we consider problems for which this knowledge is available.

\begin{figure}[tb]
   \centering
   \includegraphics[width=0.45\textwidth]{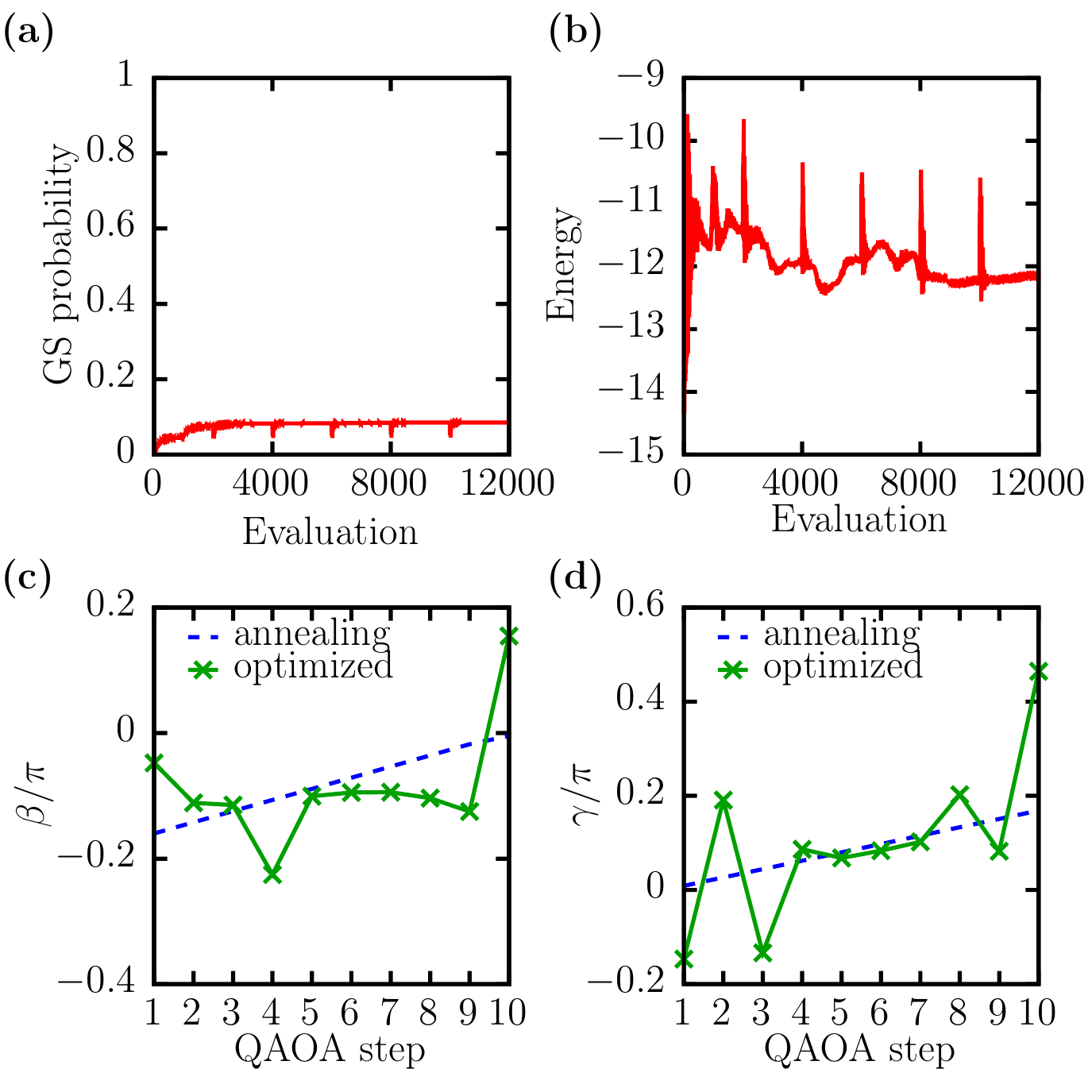}
   \caption{(color online)
   Same as Fig.~\ref{fig:2sat_18spin_p2_random}
   except that the initial values are chosen according to a linear annealing scheme with step size $\tau=0.558$ (dashed (blue) line).
   Values after optimization are marked by (green) crosses.}
   \label{fig:2sat_18spin_p2}
\end{figure}

When the function to be optimized has many local optima, the choice of the initial values
can have a strong influence on the output of the optimization algorithm.
We find that the initialization of the $\gamma_i$'s and $\beta_i$'s
seems to be crucial for the success probability that can be obtained, suggesting that there are
many local minima or stationary points.
This is illustrated in Fig.~\ref{fig:2sat_18spin_p2} where we
show the results of minimizing the success probability
starting from $\gamma_i$'s and $\beta_i$'s taken from
a linear annealing scheme (see Appendix~\ref{app:relation}),
for the same problem as the one used to produce the data
shown in Fig.~\ref{fig:2sat_18spin_p2_random}.
Looking at Fig.~\ref{fig:2sat_18spin_p2_random}(a,b), we see that
the final success probability is 38.6\% and $E_{p=10}\approx-14.22$,
whereas from Fig.~\ref{fig:2sat_18spin_p2}(a,b), we deduce that
the final success probability is only 8.5\% and $E_{p=10}\approx -12.16$.
For comparison, the actual ground state energy is $E_C^{(0)}=-19$.
Comparing also Figs.~\ref{fig:2sat_18spin_p2_random}(c,d)
and Fig.~\ref{fig:2sat_18spin_p2}(c,d)
clearly shows the impact of the initial values of
the $\gamma_i$'s and $\beta_i$'s on the results of the values
after minimization.

For this particular 18-variable 2-SAT problem,
minimizing the energy expectation $E_{p=10}$ instead of the success probability did not lead to a
higher success probability. In fact, the success probability only reached 0.1\%
and $E_{p=10}\approx-14.97$ (data not shown).
Although this energy expectation value and the initial value for the case shown in Fig.~\ref{fig:2sat_18spin_p2} ($E_{p=10}\approx-14.36$) are better than the final expectation value in the case presented in Fig.~\ref{fig:2sat_18spin_p2_random}, the success probabilities are much worse.
From these results, we conclude that the optimization of $\vec\gamma$ and $\vec\beta$ with respect to the energy expectation value may in general result in different (local) optima
than would be obtained by an optimization with respect to the success probability.
Possible reasons for this might be that the energy landscape has (many) more local minima
than the landscape of the success probability has local maxima or that the
positions of the (local) minima in the energy landscape are not aligned with (local) maxima of
the landscape of the success probability.

\begin{figure}[tb]
   \centering
   \includegraphics[width=0.45\textwidth]{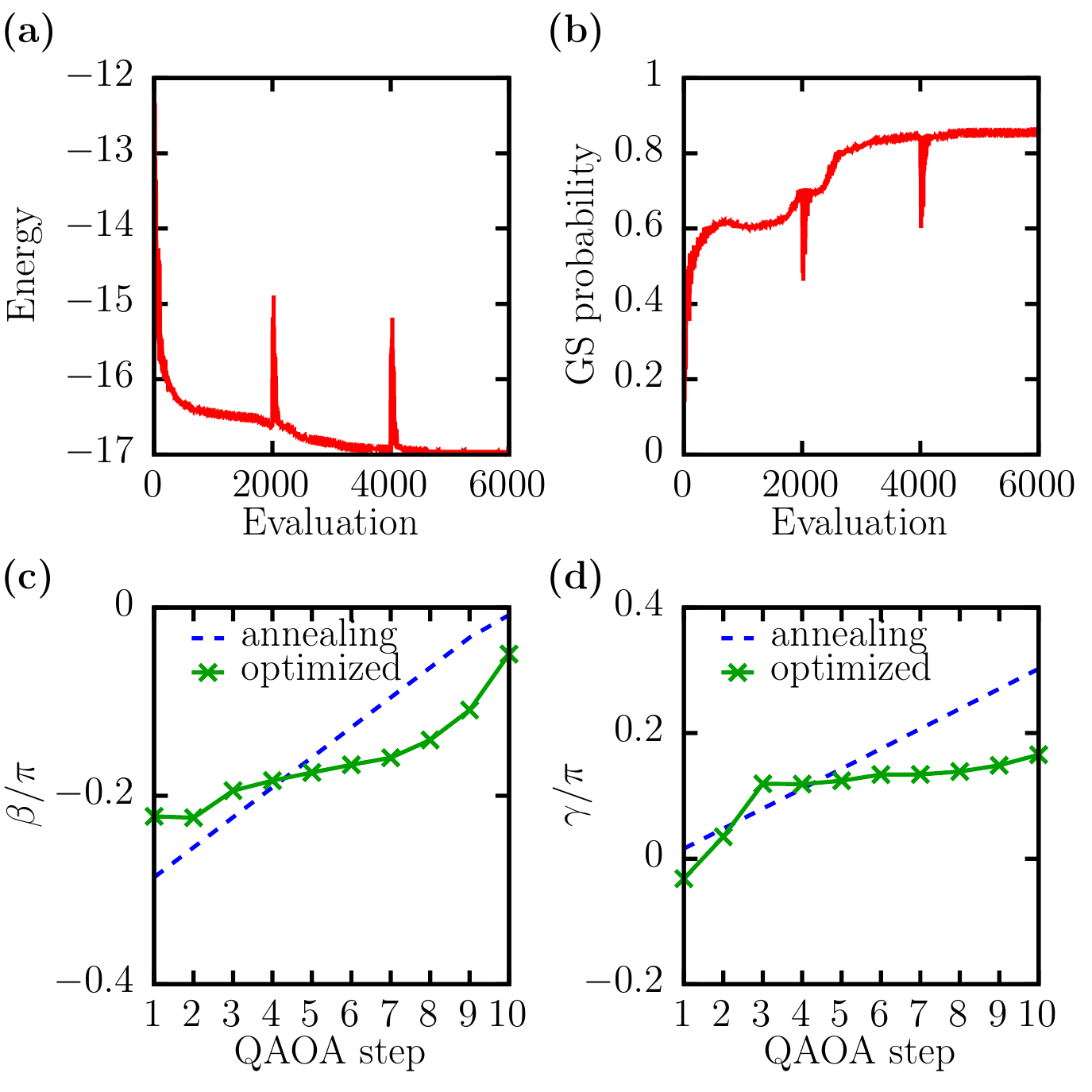}
   \caption{(color online)
   Same as Fig.~\ref{fig:2sat_18spin_p2_random}
   except that (i) the results are for the 16-variable weighted MaxCut problem instance, (ii) the initial values are chosen according to a linear annealing scheme with step size $\tau =1$ (dashed (blue) line), and
   (iii) the energy expectation value $E_{p=10}$ is taken as the cost function for the Nelder-Mead
   minimization procedure. The actual ground state energy is $E_C^{(0)}= -17.7$.
   Values after optimization are marked by (green) crosses.}
\label{fig:maxcut_16}
\end{figure}

Figure~\ref{fig:maxcut_16} shows results for a 16-variable weighted MaxCut problem
for which minimizing $E_{p=10}$ improves the success probability.
The initialization is done according to the linear annealing scheme (see Appendix~\ref{app:relation}).
This is a clear indication that for finite $p$, the QAOA can be
viewed as a tool for producing optimized annealing schemes~\cite{crooks18,zhou18}.
For this problem, the success probability after 6000 Nelder-Mead iterations is quite large ($\approx 85.6\%$).
At the end of the minimization procedure,
the $\gamma_i$'s and $\beta_i$'s deviate from their initial values (see Fig.~\ref{fig:maxcut_16}(c,d))
but, as a function of the QAOA step $i$, show the same trends, as in Fig.~\ref{fig:2sat_18spin_p2}.
This suggests that the QAOA
may yield $\gamma_i$'s and $\beta_i$'s that deviate less and less from
their values of the linear annealing scheme as $p$ increases.

\begin{figure}[tb]
   \centering \includegraphics[width=0.45\textwidth]{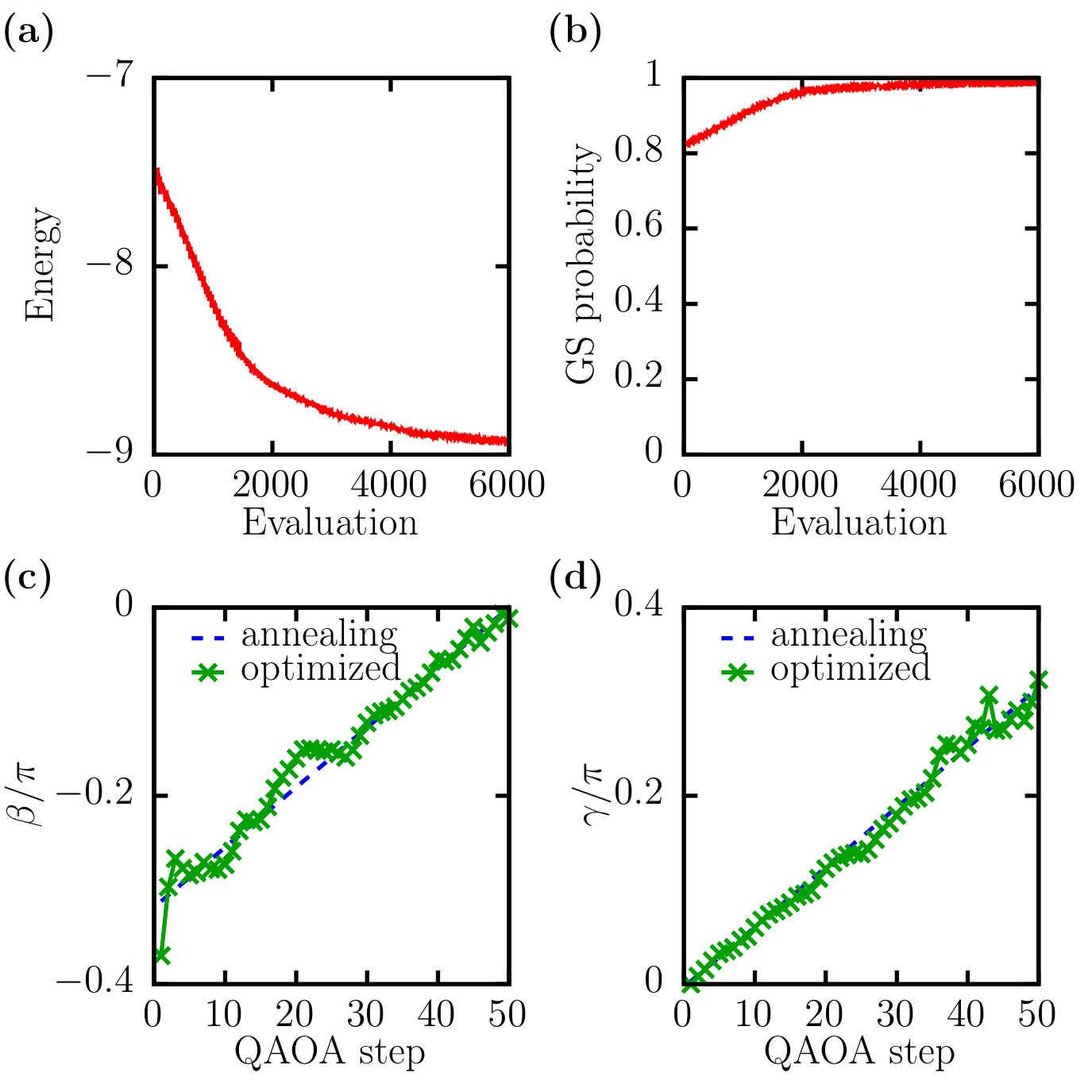}
   \caption{(color online)
   Same as Fig.~\ref{fig:2sat_18spin_p2_random}
   except that (i) the results are for the 8-variable 2-SAT problem instance (A), (ii) the initial values are chosen according to a linear annealing scheme with step size $\tau=1$ (dashed (blue) line), and (iii) the energy expectation value $E_{p=50}$ is taken as the cost function for the Nelder-Mead minimization procedure. The actual ground state energy is $E_C^{(0)}=-9$.}
\label{fig:2sat_8spin_p19}
\end{figure}

This observation is confirmed by the results shown in Fig.~\ref{fig:2sat_8spin_p19}
for an 8-variable 2-SAT problem instance.
We set $p=50$ and use the linear annealing scheme to initialize the $\gamma_i$'s and $\beta_i$'s (see Appendix~\ref{app:relation}) which yields a success probability of about 82.7\%.
Although we are using $E_{p=50}$ as the function to be minimized,
Fig.~\ref{fig:2sat_8spin_p19}(b) shows that the success probability
at the end of the minimization process is close to one.
Further optimization of the $\gamma_i$'s and $\beta_i$'s in the spirit
of the QAOA shows that small deviations of $\gamma_i$'s and $\beta_i$'s from
the linear annealing scheme increase the success probability to almost one.
Not surprisingly, this indicates that if the initial
$\gamma_i$'s and $\beta_i$'s define a trial wave function which
yields a good approximation to the ground state,
the variational approach works well~\cite{TING19}.

All in all, we conclude that the success of the QAOA strongly depends on the
problem instance. While the investigated 8-variable 2-SAT problem and the
16-variable MaxCut problem work well, the success of the (also for quantum annealing
hard) 2-SAT problem with 18 variables is rather limited.

\subsection{Quantum annealing on a D-Wave machine}

Since the QAOA results produced by a real quantum device
are of rather poor quality, for comparing the QAOA
to quantum annealing on the D-Wave quantum annealer, we eliminate all device errors
of the former by using simulators to perform the necessary quantum gate operations.

Table~\ref{tab:d_wave_data} summarizes the simulation results of
the QAOA for $p=1$ and $p=5$ in comparison with the data obtained from the
D-Wave 2000Q for 2-SAT problems with 8, 12 and 18 variables.
Both the success probability and the ratio $r$ are shown.
We present data for annealing times of $3\,\upmu\mathrm{s}$
(approximately the real time it takes the IBM Q Experience to
return one sample for the $p=1$ QAOA quantum gate circuit) and
$30\,\upmu\mathrm{s}$. Postprocessing on the D-Wave 2000Q quantum annealer has been turned off.
The results for the QAOA with $p=2,3,4,5$ steps are obtained by initializing the
$\gamma_i$'s and $\beta_i$'s ($i=1,\dots, p-1$) with the optimal values obtained from the
minimization for $p-1$ steps and setting $\gamma_p=\beta_p=0$.

\begin{table}[tb]
  \centering
  \caption{
Results produced by the D-Wave 2000Q (DW\_2000Q\_2\_1 chip) in comparison with the QAOA for various
2-SAT problem instances. Performance measures are the success probability
and the ratio $r$.
For the QAOA, parameter optimization
uses the energy expectation value $E_p$ as the function to be minimized,
as if the QAOA was executed on a real device.
The $\gamma_i$'s and $\beta_i$'s ($i=1,\dots, p-1$) are initialized with the optimal values obtained from the minimization for $p-1$ steps and $\gamma_p=\beta_p=0$.
}
\label{tab:d_wave_data}
\medskip
\begin{tabular}{c||c c|c c||c c|c c}
   Variables & \multicolumn{4}{c||}{Success probability (\%)} &\multicolumn{4}{c}{Ratio $r$}\\
   (problem  & \multicolumn{2}{c|}{D-Wave} & \multicolumn{2}{c||}{QAOA} & \multicolumn{2}{c|}{D-Wave} & \multicolumn{2}{c}{QAOA}\\
   label)   & $3\,\upmu\mathrm{s}$ & $\,30\,\upmu\mathrm{s}$ & $p=1$ & $p=5$ &$3\,\upmu\mathrm{s}$ & $\,30\,\upmu\mathrm{s}$ & $p=1$ & $p=5$\\
      \hline\hline
8  (A)  & $99.75$ & $99.92$ & $8.84$ & $42.39$ & $ 1.00$ & $ 1.00$ & $0.71$ & $0.84$\\
8  (B)  & $99.76$ & $99.92$ & $8.84$ & $42.39$ & $ 1.00$ & $ 1.00$ & $0.71$ & $0.84$\\
8  (C)  & $99.76$ & $99.92$ & $8.84$ & $42.39$ & $ 1.00$ & $ 1.00$ & $0.71$ & $0.84$\\
8  (D)  & $99.88$ & $99.95$ & $7.50$ & $46.73$ & $ 1.00$ & $ 1.00$ & $0.70$ & $0.85$\\
\hline
12  (A) & $27.90$ & $ 61.11 $  &$1.87$ & $4.08$ & $0.91$ & $0.95$ & $0.81$& $0.87$\\
12  (B) & $75.86$ & $92.48$ & $2.15$ & $7.10$ & $0.97$ & $0.99$ & $0.76$ & $0.86$\\
12  (C) & $88.12$ & $95.79$ & $2.18$ & $10.30$ & $0.98$ & $0.99$ & $0.77$ & $0.87$\\
12  (D) & $52.68$ & $77.69$ &  $2.00$ & $5.44$ & $0.93$ & $0.97$ & $0.77$ & $0.86$\\
\hline
18 (A) & $1.22$ & $4.82$ & $0.22$ & $0.87$& $0.91$ & $0.91$ & $0.79$ &$0.87$\\
18 (B) & $10.85$ & $36.12$ & $0.33$ & $1.82$ & $0.91$ & $0.94$ & $0.77$ & $0.88$\\
18 (C) & $0.88$ & $3.37$ & $0.25$ & $0.22$ & $0.91$ & $0.91$ & $0.80$ & $0.90$\\
18 (D) & $89.38$ & $97.01$ & $0.34$ & $9.03$ & $0.99$ & $ 1.00$ & $0.76$ & $0.89$\\
18 (E) & $51.20$ & $78.32$ & $0.34$ & $4.15$ & $0.95$ & $0.98$ & $0.76$ & $0.88$\\
18 (F) & $4.84$ & $16.03$ & $0.31$ & $0.83$ & $0.90$ & $0.92$ & $0.78$ & $0.89$
      \end{tabular}
\end{table}

From Table~\ref{tab:d_wave_data}, we conclude
that using the D-Wave, the probability of sampling the ground state (i.e., the unique
solution of the 2-SAT problem) is much larger than the one obtained
from running the QAOA on a simulator.
Accordingly, the ratio $r$ is also higher. However, the ratios $r$ obtained from
the D-Wave data show stronger variation with the particular
problem instance (for 12 and 18 spins) than the ratios obtained from the QAOA which seem to  systematically increase with the problem size for $p=5$. The
increase in the ratio $r$ from the QAOA for $p=1$ to the QAOA for $p=5$ is
much larger than the increase in the ratio $r$ for the D-Wave 2000Q
when using a ten times longer annealing time.
On the other hand, the ratio $r$ obtained from the D-Wave data is,
in most cases, significantly larger than the one obtained from the QAOA.
The D-Wave results and the QAOA results for $p=5$ exhibit similar trends: For many of the 12- and 18-spin problem instances, the success probabilities of the QAOA are roughly one-tenth of the probabilities obtained from the D-Wave machine for annealing times of $3\upmu$s, indicating that problem instances which are hard for the D-Wave machine are also hard for the QAOA with a small number of steps.

\section{Conclusion}\label{sec:conclusion}

We have studied the performance of the quantum approximate optimization algorithm
by applying it to a set of instances of 2-SAT problems with up to 18 variables and a unique solution,
and weighted MaxCut problems with 16 variables.

For benchmarking purposes, we only consider problems for which the solution, i.e.\
the true ground state of the problem Hamiltonian is known.
In this case, the success probability, i.e.\ the probability
to sample the true ground state, can be used as the function to be minimized.
This is the ideal setting for scrutinizing the performance of the QAOA.
In a practically relevant setting, the true ground state is not known
and one has to resort to minimizing the expectation value of the problem Hamiltonian.
Furthermore, on a real device, this expectation value needs to be estimated from a (small) sample.
Using a simulator, one can dispense of the sampling aspect.
Our simulation data show that the success of the QAOA
based on minimizing the expectation value of the problem Hamiltonian
strongly depends on the problem instance.

For a small number of QAOA steps $p=1,\ldots,50$, the QAOA may be viewed as a method to
determine the $2p$ parameters in a particular variational ansatz for the wave function.
For our whole problem collection, we find that
the effect of optimizing the $p=1$ wave function on the success probability
is rather modest, even when we run the QAOA on the simulator.
In the case of a nontrivial 8-variable 2-SAT problem,
for which the $p=1$ QAOA on a simulator yields good results,
the IBM Q Experience produced rather poor results.

There exist 2-SAT problems for which the $p=5$ QAOA performs satisfactorily
(meaning that the success probability is much larger than 1\%),
also if we perform the simulation in the practically relevant setting,
that is we minimize the expectation of the problem Hamiltonian, not the success probability.
We also observed that (local) maxima of the success probability and (local) minima of the energy expectation value seem not always to be sufficiently aligned.

Quantum annealing can be viewed as a particular realization of the QAOA with $p\to\infty$.
This suggests that we may use, for instance, a linear annealing scheme to initialize the $2p$ parameters.
For small values of $p$, after minimizing these parameters,
depending on the problem instance, they may or may not resemble the annealing scheme.
For the case with $p=50$ studied in this paper, they are close to their values of the
linear annealing scheme, yielding a success probability that is close to one.
Summarizing, the performance of the QAOA varies considerably with the problem instance,
the number of parameters $2p$, and their initialization.
This variation also makes it difficult to develop a general strategy for optimizing the $2p$ parameters.

For the set of problem instances considered, taking the success probability as a measure,
the QAOA cannot compete with quantum annealing when no minor embedding is necessary (as in the case of the instances studied).
We also find a correlation between instances that are hard for quantum annealing and instances that are hard for the QAOA.
The ratio $r$, which also requires knowledge of the true ground state,
is a less sensitive measure for the algorithm performance.
Therefore, it shows less variation from one problem instance to another.
But the ratios $r$ obtained from the QAOA (using a simulator) are, with a few exceptions, still
significantly smaller than those obtained by quantum annealing on a real device.

\begin{acknowledgments}
Access and compute time on the D-Wave machine located at the headquarters of D-Wave Systems in Burnaby (Canada) were provided by D-Wave Systems.
We acknowledge use of the IBM Q Experience. This work does not reflect the views or opinions of IBM or any of its employees.
D.W.~is supported by the Initiative and Networking Fund of the Helmholtz Association through the Strategic Future Field of Research project ``Scalable solid state quantum computing (ZT-0013)''.
K.M.~acknowledges support from the project OpenSuperQ (820363) of the EU Flagship Quantum Technologies.
\end{acknowledgments}

\appendix

\section{Relation between QAOA and quantum annealing}\label{app:relation}
In this appendix, we give a mapping between a Hamiltonian describing a quantum annealing scheme and the QAOA for a given number of steps. The annealing Hamiltonian reads
\begin{align}
   H(s) = A(s)(-H_0) + B(s)H_C,\quad s=t/t_a\in[0,1],
\end{align}
where (we have to add an additional minus sign to $H_0$ such that the state $\ket{+}^{\otimes N}$ we start from is the ground state of $H(s)$ and the convention still conforms with the formulation of the QAOA)
\begin{align}
   H_0 &= \sum_i \sigma_i^x,\\
   H_C &= \sum_i h_i\sigma_i^z +\sum_{ij}J_{ij}\sigma_i^z\sigma_j^z.
\end{align}
We discretize the time-evolution operator of the annealing process into $N$ time steps of size $\tau=t_a/N$. Approximating each time step to second order in $\tau$ yields~\cite{deraedt83,suzuki85}
\begin{align}
   U &= e^{+i\tau A(s_{N})H_0/2} e^{-i\tau B(s_{N})H_C}\nonumber\\
   &\quad\times e^{+i\tau (A(s_{N})+A(s_{N-1}))H_0/2}\cdots\nonumber\nonumber\\
   &\quad\times e^{-i\tau B(s_{2})H_C} e^{+i\tau (A(s_{2})+A(s_{1}))H_0/2}\nonumber\\
   &\quad\times e^{-i\tau B(s_{1})H_C} e^{+i\tau A(s_{1})H_0/2}\label{U_qa},
\end{align}
where $s_n = (n-1/2)/N$, and $n=1,\dots, N$.

To map Eq.~(\ref{U_qa}) to the QAOA evolution
\begin{align}
   V=e^{-i\beta_pH_0}e^{-i\gamma_pH_C}\cdots e^{-i\beta_1H_0}e^{-i\gamma_1H_C},
\end{align}
we can neglect $e^{+i\tau A(s_1)H_0/2}$ because its action on $\ket{+}^{\otimes N}$ yields only a global phase factor and we can choose
\begin{align}
   \gamma_n &= \tau B(s_n),\qquad n = 1,\dots ,N \label{eq_gamma}\\
   \beta_n &= -\tau \left( A(s_{n+1})+A(s_{n}) \right)/2,\qquad n = 1,\dots ,N-1\\
   \beta_{N} &= -\tau A(s_N)/2.\label{eq_beta}
\end{align}
So $N$ time steps for the second-order-accurate annealing scheme correspond to $p=N$ steps for the QAOA.

As an example, we take
\begin{align}
   A(s) = 1-s,\qquad B(s) = s.
\end{align}
Using Eqs.~(\ref{eq_gamma}) -- (\ref{eq_beta}), we obtain
\begin{align}
   \gamma_n &= \frac{\tau(n-1/2)}{N}\\
   \beta_n &= -\tau\left(1-\frac{n}{N}\right)\\
   \beta_N &= -\frac{\tau}{4N}.
\end{align}

\section{Problem instances}\label{app:problem_instances}
\begin{table}[thb!]
  \centering
  \caption{The 16-variable weighted MaxCut problem instance.}
\label{tab:maxcut}
\smallskip
 \begin{minipage}[t]{0.15\textwidth}
   \vspace{0pt}
  \begin{tabular}{c|c||c}
   $i$ & $j$ & $J_{ij}$\\
      \hline\hline
   0 & 4 & 0.4 \\
   0 & 5 & 0.8 \\
   0 & 6 & 0.2 \\
   1 & 4 & 0.7 \\
   1 & 5 & 0.5 \\
   1 & 6 & 0.6 \\
   1 & 7 & 0.8 \\
   2 & 4 & 0.4 \\
   2 & 5 & 1.0 \\
   2 & 6 & 0.3 \\
   2 & 7 & 0.7
   \end{tabular}
   \end{minipage}
    \begin{minipage}[t]{0.15\textwidth}
   \vspace{0pt}
      \begin{tabular}{c|c||c}
   $i$ & $j$ & $J_{ij}$\\
      \hline\hline
   3 & 4 & 0.3 \\
   3 & 5 & 0.7 \\
   3 & 6 & 0.6 \\
   3 & 7 & 0.4 \\
   4 & 12 & 0.1 \\
   6 & 14 & 0.2 \\
   7 & 15 & 1.0 \\
   8 & 12 & 0.1 \\
   8 & 13 & 0.9 \\
   8 & 14 & 1.0 \\
   8 & 15 & 0.8
       \end{tabular}
   \end{minipage}
   \begin{minipage}[t]{0.15\textwidth}
   \vspace{0pt}
      \begin{tabular}{c|c||c}
   $i$ & $j$ & $J_{ij}$\\
      \hline\hline
   9 & 12 & 0.3 \\
   9 & 13 & 0.5 \\
   9 & 14 & 0.1 \\
   9 & 15 & 0.7 \\
   10 & 12 & 0.5 \\
   10 & 13 & 0.7 \\
   10 & 14 & 0.3 \\
   10 & 15 & 0.6 \\
   11 & 12 & 0.2 \\
   11 & 13 & 0.8 \\
   11 & 14 & 0.5
       \end{tabular}
   \end{minipage}
\end{table}
The problem instance of the 16-variable weighted MaxCut problem is listed in Table~\ref{tab:maxcut}. 

Our 2-SAT problems have been selected such that they possess a unique ground state and a highly degenerate first-excited state, making them (very) hard to solve by simulated annealing.
In this paper, we have taken instances from this collection that (1) present different degrees of difficulty for quantum annealing and (2) can be mapped directly onto the architecture of the IBM Q Melbourne chip and the Chimera graph architecture of the D-Wave 2000Q quantum annealer.
We require (2) because otherwise, we would need to perform additional swap gates on the IBM Q Experience and use a minor embedding on the D-Wave 2000Q quantum annealer.
This would make a direct comparison complicated and require including the particular graph structure in the benchmark, rendering it device-dependent and thus losing generality.
The simulator, on the other hand, does not impose any constraints on the connectivity.
Tables~\ref{tab:2sat8} -- \ref{tab:2sat18} contain the instances of the 8-, 12- and 18-variable 2-SAT problems, respectively. Entries for which both $J_{ij}$ and $h_i$ are zero have been omitted.

\begin{table*}[thbp]
  \centering
  \caption{8-variable 2-SAT problem instances with 9 clauses.}
\label{tab:2sat8}
\smallskip
\begin{minipage}{0.22\textwidth}
 \vspace{0pt}
  \begin{tabular}{c|c||c|c}
   \multicolumn{4}{c}{(A)} \\
   $i$  & $j$ & $J_{ij}$ & $h_{i}$ \\
      \hline\hline
   0 & 6 & 1 & 0\\
   1 & 3 & $-1$ & 0\\
   2 & 0 & $-1$ & 1 \\
   3 & 4 & 1 & 0\\
   5 & 6 & $-1$ & $-1$\\
   6 & 4 & $-1$ & 1\\
   7 & 1 & $-1$ & 1
  \end{tabular}
\end{minipage}
\begin{minipage}{0.22\textwidth}
 \vspace{0pt}
  \begin{tabular}{c|c||c|c}
   \multicolumn{4}{c}{(B)}\\
   $i$ & $j$ & $J_{ij}$ & $h_{i}$ \\
      \hline\hline
   0 & 7 & $-1$ & $-1$\\
   1 & 3 & $-1$ & 1\\
   2 & 5 & $1$ & 0\\
   2 & 6 & $-1$ & 0\\
   4 & 6 & $-1$ & $-1$\\
   5 & 7 & $1$ & 0\\
   6 & 3 & $1$ & $1$
  \end{tabular}
\end{minipage}
\begin{minipage}{0.22\textwidth}
 \vspace{0pt}
  \begin{tabular}{c|c||c|c}
   \multicolumn{4}{c}{(C)}\\
   $i$ & $j$ & $J_{ij}$ & $h_{i}$ \\
      \hline\hline
   0 & 3 & $-1$ & $-1$\\
   0 & 7 & $-1$ & $-1$\\
   1 & 2 & 1 & 0\\
   2 & 3 & $1$ & 0\\
   4 & 0 & $1$ & $-1$ \\
   5 & 1 & $1$ & $-1$ \\
   6 & 7 & $-1$ & 1
  \end{tabular}
\end{minipage}
\begin{minipage}{0.22\textwidth}
   \vspace{0pt}
  \begin{tabular}{c|c||c|c}
   \multicolumn{4}{c}{(D)}\\
   $i$ & $j$ & $J_{ij}$ & $h_{i}$ \\
      \hline\hline
   0 & 2 & $-1$ & 0\\
   0 & 7 & $1$ & 0\\
   1 & 6 & $1$ & 0\\
   1 & 7 & $-1$ & 0\\
   2 & 3 & $1$  & 0\\
   3 & 4 & $1$ & $-1$\\
   3 & 6 & $-1$ & $-1$\\
   4 & 5 & 1 & 0\\
   6 & 5 & 1 & 1
  \end{tabular}
\end{minipage}
\end{table*}

\begin{table*}[thbp]
  \centering
  \caption{12-variable 2-SAT problem instances with 13 clauses.}
\label{tab:2sat12}
\smallskip
\begin{minipage}{0.22\textwidth}
 \vspace{0pt}
  \begin{tabular}{c|c||c|c}
   \multicolumn{4}{c}{(A)}\\
   $i$ & $j$ & $J_{ij}$ & $h_{i}$ \\
      \hline\hline
   0 & 9 & 1 & $-1$\\
   1 & 9 & $-1$  & 1\\
   2 & 6 & $1$  & 1\\
   3 & 11 & 1 & 1\\
   5 & 11 & $1$ & 1\\
   6 & 10 & $-1$ & 1 \\
   7 & 6 & $-1$ & $-1$\\
   8 & 10 & $-1$ & $-1$ \\
   9 & 4 & $1$ & $-2$ \\
   10 & 4 & $-1$ & 1\\
   11 & 9 & 1 & $1$
  \end{tabular}
\end{minipage}
\begin{minipage}{0.22\textwidth}
 \vspace{0pt}
  \begin{tabular}{c|c||c|c}
   \multicolumn{4}{c}{(B)}\\
   $i$ & $j$ & $J_{ij}$ & $h_{i}$ \\
      \hline\hline
   0 & 7 & $1$ & $-1$\\
   0 & 8 & $1$ & $-1$\\
   1 & 10 & $1$ & $-1$\\
   2 & 4 & $-1$ & $-2$\\
   2 & 8 & $1$ & $-2$\\
   3 & 0 & $1$ & $-1$ \\
   5 & 4 & 1 & $-1$ \\
   6 & 2 & $-1$ & 1 \\
   7 & 11 & 1 & 0\\
   9 & 2 & $-1$ & 1 \\
   10 & 11 & $-1$ & 0
  \end{tabular}
\end{minipage}
\begin{minipage}{0.22\textwidth}
 \vspace{0pt}
  \begin{tabular}{c|c||c|c}
   \multicolumn{4}{c}{(C)}\\
   $i$ & $j$ & $J_{ij}$ & $h_{i}$ \\
      \hline\hline
   0 & 6 & $1$ & 0\\
   1 & 5 & 1 & $-1$\\
   2 & 0 & $-1$ & $-1$ \\
   3 & 10 & $-1$ & 0\\
   4 & 11 & $-1$ & $-1$ \\
   5 & 3 & $-1$ & $-1$ \\
   6 & 7 &  1  & $-1$\\
   7 & 10 & 1 & 0\\
   8 & 5 & $-1$ & $1$\\
   9 & 11 & 1 & 1 \\
   11 & 6 & 1 & 1
  \end{tabular}
\end{minipage}
\begin{minipage}{0.22\textwidth}
   \vspace{0pt}
  \begin{tabular}{c|c||c|c}
   \multicolumn{4}{c}{(D)}\\
   $i$ & $j$ & $J_{ij}$ & $h_{i}$ \\
      \hline\hline
   0 & 10 & $-1$ & 1\\
   1 & 9 & $-1$ & 0\\
   2 & 10 & $-1$ & 1\\
   3 & 6 & $-1$ & $-1$\\
   4 & 3 & $-1$ & 1\\
   5 & 10 & $-1$ & 1\\
   6 & 9 & $-1$ & 0\\
   7 & 1 & $1$ & $-1$ \\
   8 & 3 & $-1$ & $-1$ \\
   10 & 8 & $-1$ & $-2$\\
   11 & 8 & 1 & $-1$
  \end{tabular}
\end{minipage}
\end{table*}

\begin{table*}[thbp]
  \centering
  \caption{18-variable 2-SAT problem instances with 19 clauses.}
\label{tab:2sat18}
\smallskip
\begin{minipage}{0.16\textwidth}
 \vspace{0pt}
  \begin{tabular}{c|c||c|c}
   \multicolumn{4}{c}{(A)}\\
   $i$ & $j$ & $J_{ij}$ & $h_{i}$ \\
      \hline\hline
   0 & 5 & $-1$ & $-2$\\
   0 & 7 & 1 & $-2$\\
   0 & 10 & $-1$ & $-2$\\
   1 & 13 & 1  & $-2$\\
   1 & 15 & 1 & $-2$\\
   2 & 5 & $-1$ & 1\\
   3 & 12 & $-1$ & 1\\
   4 & 7 & 1  & 1\\
   6 & 11 & 1 & 1\\
   7 & 9 & 1 & 1\\
   8 & 0 & $-1$ & 1 \\
   9 & 12 & $-1$ & 0\\
   10 & 11 & 1 & 0\\
   11 & 15 & $-1$ & 1 \\
   14 & 13 & $-1$ & $-1$\\
   16 & 1 & $-1$ & 1 \\
   17 & 1 & 1 & $-1$\\
   \end{tabular}
   \end{minipage}
   \begin{minipage}{0.16\textwidth}
 \vspace{0pt}
  \begin{tabular}{c|c||c|c}
   \multicolumn{4}{c}{(B)}\\
   $i$ & $j$ & $J_{ij}$ & $h_{i}$ \\
      \hline\hline
   0 & 1 & 1 & $-1$\\
   0 & 3 & 1 & $-1$\\
   0 & 11 & $-1$ & $-1$\\
   1 & 13 & 1 & 2\\
   1 & 15 & 1 & 2\\
   1 & 17 & $-1$ & 2\\
   2 & 10 & 1 & 0\\
   3 & 12 & 1 & 0\\
   4 & 13 & 1 & $-1$\\
   5 & 8 & $-1$ & 0\\
   5 & 12 & 1 & 0\\
   6 & 17 & 1 & 1\\
   7 & 11 & 1  & 0\\
   8 & 10 & 1 & 0\\
   9 & 15 & $-1$ & 1 \\
   14 & 7 & $-1$ & $-1$\\
   16 & 2 & 1 & 1
  \end{tabular}
  \end{minipage}
  \begin{minipage}{0.16\textwidth}
  \vspace{0pt}
  \begin{tabular}{c|c||c|c}
   \multicolumn{4}{c}{(C)}\\
   $i$ & $j$ & $J_{ij}$ & $h_{i}$ \\
      \hline\hline
   1 & 0 & $-1$ & $-1$ \\
   2 & 4 & $-1$ & 0\\
   3 & 4 & $-1$ & 1 \\
   4 & 0 & $-1$ & $-2$ \\
   5 & 4 & $-1$ & $-1$ \\
   6 & 1 & $-1$ & 1 \\
   7 & 2 & $-1$ & 1 \\
   8 & 1 & 1 & $-1$\\
   9 & 10 & 1 & 1\\
   10 & 12 & 1 & 1\\
   11 & 16 & $-1$ & $-1$\\
   12 & 15 & $-1$ & 0\\
   13 & 10 & $-1$ & $-1$\\
   14 & 15 & $1$ & $-1$\\
   15 & 16 & 1 & $-1$ \\
   16 & 5 & 1 & 1 \\
   17 & 5 & $-1$ & 1
  \end{tabular}
  \end{minipage}
  \begin{minipage}{0.16\textwidth}
  \vspace{0pt}
  \begin{tabular}{c|c||c|c}
   \multicolumn{4}{c}{(D)}\\
   $i$ & $j$ & $J_{ij}$ & $h_{i}$ \\
      \hline\hline
   0 & 9 & $1$ & 0\\
   0 & 14 & $1$ & 0\\
   1 & 4 & $-1$ & $-1$\\
   2 & 10 & $-1$ & 0\\
   2 & 13 & $-1$ & 0\\
   3 & 17 & $1$ & 1\\
   4 & 6 & $-1$  & 1\\
   4 & 7 & $-1$ & 1\\
   5 & 10 & 1 & 0\\
   5 & 11 & 1 & 0\\
   6 & 15 & $1$ & 0 \\
   7 & 9 & $1$ & 0\\
   8 & 15 & $1$ & 0\\
   8 & 17 & $1$ & 0\\
   11 & 14 & $-1$ & 0\\
   12 & 16 & $-1$ & 1\\
   13 & 16 & $-1$ & 0
  \end{tabular}\\
  \end{minipage}
  \begin{minipage}{0.16\textwidth}
  \vspace{0pt}
  \begin{tabular}{c|c||c|c}
   \multicolumn{4}{c}{(E)}\\
   $i$ & $j$ & $J_{ij}$ & $h_{i}$ \\
      \hline\hline
   0 & 5 & $1$ & 0\\
   0 & 9 & $-1$ & 0\\
   1 & 6 & $1$ & 0\\
   1 & 14 & $1$ & 0\\
   2 & 10 & $1$ & $-1$ \\
   3 & 5 & $-1$ & 0\\
   3 & 15 & $1$ & 0\\
   4 & 17 & $1$ & 0\\
   4 & 8 & $-1$ & 0\\
   6 & 15 & $-1$ & 0 \\
   7 & 9 & $-1$ & $-1$\\
   8 & 13 & $-1$ & 0\\
   10 & 17 & $-1$ & 0\\
   11 & 7 & $1$ & $-1$\\
   12 & 14 & $-1$ & 1\\
   16 & 13 & $-1$ & $-1$ \\
   17 & 7 & $-1$ & $-1$
  \end{tabular}\\
  \end{minipage}
  \begin{minipage}{0.16\textwidth}
  \vspace{0pt}
  \begin{tabular}{c|c||c|c}
   \multicolumn{4}{c}{(F)}\\
   $i$ & $j$ & $J_{ij}$ & $h_{i}$ \\
      \hline\hline
   0 & 3 & $-1$ & $-1$ \\
   0 & 4 & $1$ & $-1$\\
   1 & 6 & $1$ & 0\\
   2 & 13 & $-1$ & $-1$\\
   3 & 1 & $-1$ & $-1$ \\
   4 & 9 & $1$ & 1\\
   5 & 4 & $-1$ & $-1$\\
   6 & 14 & $-1$ & 0 \\
   7 & 0 & $-1$ & 1 \\
   8 & 3 & $1$ & $-1$ \\
   9 & 10 & $-1$ & $-1$\\
   9 & 15 & $1$ & $-1$\\
   11 & 10 & $1$ & $-1$\\
   13 & 14 & $-1$ & 0\\
   15 & 12 & $1$ & 1 \\
   16 & 12 & $-1$ & $1$ \\
   17 & 15 & $-1$ & $-1$
  \end{tabular}\\
  \end{minipage}
\end{table*}

\bibliographystyle{apsrev4-1}
\bibliography{bibliography_qaoa}

\end{document}